\documentclass[aps,pre,twocolumn]{revtex4}

\usepackage{latexsym}
\usepackage{amsmath}
\usepackage{amsfonts}
\usepackage{amssymb}
\usepackage{amsbsy}
\usepackage{bm}
\usepackage{color}
\usepackage{graphicx}
\usepackage{float}
\usepackage{enumitem}

\newcommand{\lla}{\left\langle}
\newcommand{\rra}{\right\rangle}

\newcommand{\brh}{\mathrm{\bf H}}

\newcommand{\vphi}{\varphi}
\newcommand{\ttau}{\tilde \tau}
\newcommand{\ttone}{\tilde \tau_{1}}
\newcommand{\lbar}[1]{\overline{#1}}
\newcommand{\dst}{\displaystyle}

\newcommand{\rev}[1]{\textcolor{black}{#1}}

\graphicspath{{./figures/}}

\begin{document}
\begin{titlepage}
\title{Hydrodynamics of polymers in an active bath }
\author{Aitor Martin-Gomez, Thomas Eisenstecken, Gerhard Gompper, and Roland G. Winkler}
\affiliation{Theoretical Soft Matter and Biophysics, Institute of
Complex Systems and Institute for Advanced Simulation,
Forschungszentrum J\"ulich, 52425 J\"ulich, Germany }

\begin{abstract}

The conformational and dynamical properties of active polymers in solution are determined by the nature of the activity, and the behavior of polymers with self-propelled, active Brownian particle-type monomers differs qualitatively from that of polymers with monomers driven externally by colored noise forces. We present simulation and theoretical results for polymers in solution in the presence of external active noise. In simulations, a semiflexible bead-spring chain is considered, in analytical calculations, a continuous linear wormlike chain. Activity is taken into account by independent monomer/site velocities, with orientations changing in a diffusive manner. In simulations, hydrodynamic interactions (HI) are taken into account  by the Rotne–Prager–Yamakawa tensor, or by an implementation of the active polymer in the multiparticle collision dynamics approach for fluids. To arrive at an analytical solution, the preaveraged Oseen tensor is employed. The active process implies a dependence of the stationary-state properties on HI via the polymer relaxation times. With increasing activity, HI lead to an enhanced swelling of flexible polymers, and the conformational properties differ substantially from those of polymers with self-propelled monomers in presence of HI or free-draining polymers. The polymer mean square displacement is enhanced by HI. Over a wide range of time scales, hydrodynamics leads to a subdiffusive regime of the site mean square displacement for flexible active polymers, with an exponent of $5/7$, larger than that of the Rouse ($1/2$) and Zimm ($2/3$) models of passive polymers.

\end{abstract}

\maketitle
\end{titlepage}

\section{Introduction}

Active matter is characterized by a continuous energy consumption of its agents from internal or external sources, which can be converted into  directed motion \cite{elge:15}.
The associated out-of-equilibrium nature of active matter is the origin of  fascinating phenomena, such as activity-driven phase separation or large-scale collective motion, aspects absent in corresponding passive systems \cite{marc:13,cate:15,elge:15,bech:16,rama:19,gomp:20}. A simple and generic model for a dry-active-matter agent \cite{shae:20} is the active Brownian particle (ABP), a hard-sphere- or hard-disc-type particle propelled in a body-fixed direction, which changes in a diffusive manner \cite{bech:16,roma:12,bial:12,redn:13,fily:14,cate:15,wyso:14,sten:14,elge:15,wyso:16,digr:18,das:18.1}. Computer simulations of ABP ensembles reveal motility-induced phase separation (MIPS) \cite{bech:16,bial:12,redn:13,fily:14,cate:15,wyso:14,sten:14,wyso:16,digr:18}, enhanced wall accumulation \cite{elge:13.1,fily:14,das:18.1},  and an active pressure (denoted as swim pressure) \cite{taka:14,solo:15,wink:15,fily:18,das:19}. Additional fascinating structural and dynamical properties can be expected from more complex assemblies of active particles, such as dumbbells \cite{thak:12,suma:14,furu:14,wink:16,sieb:17,gibb:17,petr:18}, linear polymers \cite{loi:11,kais:14,hard:14,ghos:14,chel:14,sark:14,jian:14.1,lask:15,shin:15,kais:15,isel:15,sama:16,eise:16,smre:17,eise:17,wink:17,mart:18.1,bian:18,loew:18,anan:18}, or more complex arrangements \cite{kuch:16}. The coupling of activity and internal degrees of freedom gives rise to novel phenomena, such as an activity-induced polymer collapse, typical in two dimensions, \cite{hard:14,kais:14,bian:18} or swelling \cite{ghos:14,shin:15,hard:14,eise:16,eise:17,mart:18.1,mous:19}, and a polymer-length-dependent suppression of phase separation \cite{suma:14,sieb:17}. This illustrates that active soft matter is a promising new class of materials with many as yet unexplored features \cite{cate:11,need:17}.

Nature provides a wide spectrum of systems, where  properties are governed by the activity  of filamentous, polymer-like building blocks and structures. Linear polymers, such as filamentous actin or microtubules of the cell cytoskeleton are propelled by tread-milling and motor proteins \cite{ridl:03,juel:07,marc:13,pros:15,cord:14,gang:12,ravi:17}. Similarly, in motility assays, filaments are propelled on carpets of motor proteins anchored on a substrate \cite{hara:87,nedl:97,scha:10,sumi:12}.  Moreover, the active dynamics of microtubules \cite{bran:09} or actin-filaments \cite{webe:15} enhances the dynamics of chromosomal loci \cite{webe:12,jave:13} and chromatin \cite{zido:13}. A characteristic feature of biological cells is the intrinsic mixture of active and passive components; specifically the active cytoskeleton and a large variety of passive colloidal and polymeric objects \cite{bran:08,mart:18.1}. Due to an accelerated dynamics of the stirred fluid in the cytoskeleton, a large variety of embedded objects, such as vesicles, passive colloids, polymeric structures, experience an enhanced stochastic motion, which implies an  enhanced random motion of tracer particles.  Similarly, countless ATP-dependent enzymatic activity-induced mechanical fluctuations drive molecular motion in the bacterial cytoplasm and the nucleus of eukaryotic cells \cite{webe:12}. Moreover,  self-propelled long swarming bacteria such as {\em Proteus mirabilis} in biofilms \cite{cope:09}  appear as semiflexible polymers, and  rodlike  objects are formed via self-assembly, e.g., by dinoflagellates \cite{sela:11,sohn:11}.

Synthetic active or activated colloidal polymers \cite{loew:18} are nowadays synthesized in various ways.
Assembly of active chains of metal-dielectric Janus colloids (monomers) can be achieved by imbalanced interactions, where simultaneously the motility and the colloid interactions are controlled by an AC electric field \cite{yan:16,dile:16,nish:18}. Electrohydrodynamic convection  rolls lead to self-assembled colloidal chains in a nematic liquid crystal matrix and directed movement \cite{sasa:14}. Moreover, chains of linked colloids, which are uniformly coated with catalytic nanoparticles, have been synthesizes \cite{bisw:17}. Hydrogen peroxide decomposition on the surfaces of the colloidal monomers generates phoretic flows, and active hydrodynamic interactions between monomers results in an enhanced diffusive motion \cite{bisw:17}.

Hydrodynamic interactions (HI) play a major role for the conformational and dynamical properties of active polymers. As has been shown in simulations, the hydrodynamic coupling between two polar externally-driven filaments   leads to cooperative effects \cite{jian:14.1}. Polymers composed of self-propelled  ABPs shrink substantially in the presence of HI at moderate activities and swell at high activity \cite{mart:19}, however, to a far less extent than dry active polymers \cite{eise:16,eise:17}.

In this article, we explore the effect of external colored noise, mimicking an active environment, on the properties of semiflexible polymers in dilute solution by hydrodynamic simulations and analytical theory. These kind of active polymers are different from polymers   with self-propelled monomers \cite{mart:19,loew:18}, as the active contribution is not force free, active forces rather give rise to Stokeslet flows. We analyze the influence of the additional hydrodynamic flow field  on the conformational and dynamical polymer properties, in comparison to self-propelled polymers. We like to emphasize that in absence of HI, the properties of active  polymers with externally-driven and self-propelled monomers are identical \cite{eise:16}.

\begin{figure}[t]
	\begin{center}
		\includegraphics[width=\columnwidth]{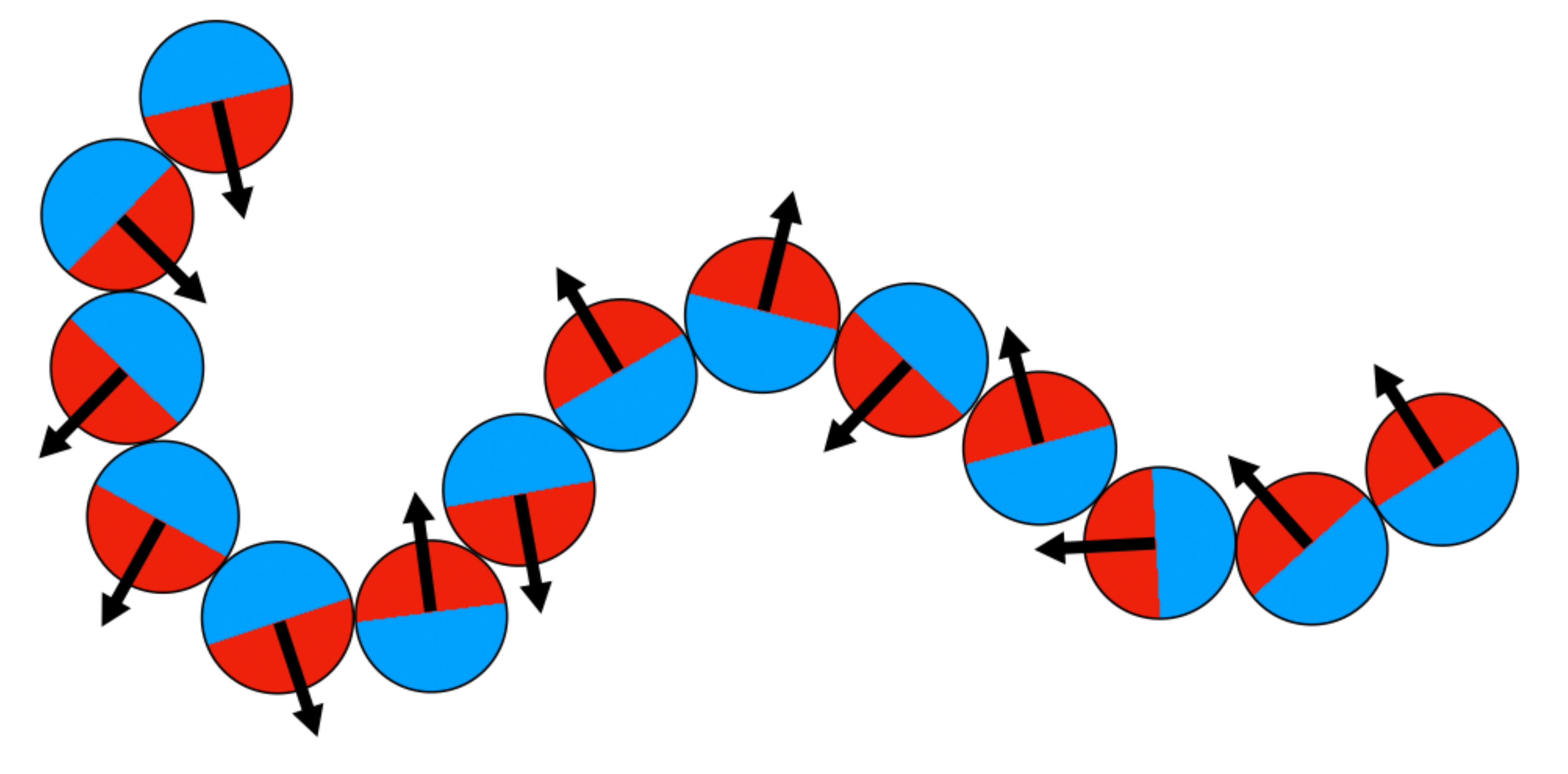}
	\end{center}\caption{Illustration of a  active polymer.  ABP monomers are drive by persistent forces (indicated by arrows) whose temporal orientation-correlations  decay exponentially.} \label{fig:sketch}
\end{figure}

In simulations, the polymers are described as  bead-spring linear phantom or self-avoiding chains with ABP-type monomers (cf. Fig.~\ref{fig:sketch}), which change their  propulsion direction in a diffusive manner \cite{eise:16}. Hydrodynamic interactions are taken into account through the Rotne-Prager-Yamakawa hydrodynamic tensor \cite{rotn:69,yama:70}. Alternatively, the same polymers are embedded in a multiparticle-collision dynamics fluid \cite{kapr:08,gomp:09}. The Gaussian semiflexible polymer model is adopted for the analytical considerations \cite{wink:94,harn:96,eise:16}, with active sites modeled by an Ornstein-Uhlenbeck process (active Ornstein-Uhlenbeck particle, AOUP) \cite{fodo:16,das:18.1,eise:16}, where the active velocity vector  changes in a diffusive manner; here, HI is included via the preaveraged Oseen tensor \cite{doi:86}. Monomer Stokeslets arise from bond, bending, and excluded-volume interactions between monomers,  thermal forces, and, in particular, from  active forces.  Hence, we capture the long-range character of HI in polymers of a broad class of externally-driven active monomers. 

Our studies reveal a decisive influence of hydrodynamic interactions on the active polymer conformations and dynamics.  Externally-driven flexible polymers monotonically swell with increasing activity, in contrast to polymers with self-propelled monomers \cite{mart:19}.   Semiflexible polymers shrink at moderated activities and swell for high activities. In the asymptotic limit of large activities, the same stretching as for free-draining active polymers is assumed. The reason is the violation of the fluctuation-dissipation theorem of the active processes, which leads to the dependence of stationary-state properties on the hydrodynamically modified relaxation times, however, in a different and less dominate way as for polymers of self-propelled monomers. The particular conformations are a consequence of the time-scale separation between the thermal process, dominating for zero or very weak activities, and the active process with hydrodynamically slowed-down relaxation times. The activity-dependent relaxation times also affect the translation motion, and a subdiffusive time regime appears, where the monomer mean square displacement (MSD) in the polymer center-of-mass reference frame exhibits a power-law dependence with the exponent $\gamma'= 5/7$, larger  than the Zimm value, $\gamma' =2/3$, of a passive polymer.

The manuscript is organized as follows. Section~\ref{sec:model-sim} describes the discrete model of the active polymer along with the simulation approaches. The results of the simulations are presented in Sec.~\ref{sec:results_sim}.  The analytical approach is introduced in Sec.~\ref{sec:model-analy}. Analytical results for the conformational and dynamical properties are discussed in Sec.~\ref{sec:conformations} and Sec.~\ref{sec:cont_dyn}, respectively.  Finally, Sec.~\ref{sec:summary} summarizes our findings. Appendix~\ref{app:lagpar} provides asymptotic results for the stretching coefficient.

\section{Computer Simulations} \label{sec:model-sim}

\subsection{Model of active polymers} \label{sec:active_polymer}

The semiflexible polymers are composed of $N_m$ active Brownian particles ($i=1,\ldots,N_m$), which are linearly connected by a harmonic bond potential, $U_l$, and experience bond-orientational restrictions by the bending potential $U_b$. Excluded-volume interactions are taken into account by the purely repulsive Lennard-Jones potential $U_{LJ}$. Explicitly, the potentials are \cite{mart:19}
\begin{align} \label{eq:pot_bond}
U_{l} = & \ \frac{\kappa_{l}}{2} \sum^{N_m}_{i=2} \left(|\bm{R}_i|  - \ l \right)^2 , \\ \label{eq:pot_bend}
U_{b} = & \ \frac{\kappa_{b}}{2} \sum^{N_m -1}_{i=2} \left(   \bm{R}_{i+1} -  \bm{R}_{i}  \right)^2 , \\ \label{eq:pot_lj}
U_{LJ} = & \
\left\{ \begin{matrix} 4 \epsilon \dst \sum_{i<j} \left[
\left( \dst \frac{\dst \sigma}{\dst  r_{ij}} \right)^{12} - \left( \dst \frac{\dst \sigma}{\dst r_{ij}} \right)^6 + \dst \frac{1}{4}\right], & r_{ij} < \sqrt[6]{2}\sigma \\
0, &r_{ij} > \sqrt[6]{2} \sigma
\end{matrix}
\right.  .
\end{align}
The coefficients $\kappa_{l}$ and $\kappa_{b}$ are the bond and bending constants, respectively, and $l$ is the equilibrium length of  the bond vector $\bm R_{i+1} = \bm r_{i+1} - \bm r_i$.  The  vector $\bm{r}_{ij}=\bm{r}_{i}-\bm{r}_{j}$ is the vector between monomers $i$ and $j$, and $r_{ij} = |\bm r_{ij}|$. The energy $\epsilon$ measures the strength of the repulsive potential and $\sigma$ defines  the particle diameter. In addition, every monomer experience an active force
\begin{align} \label{eq:active_force}
\bm F_i^a = F^a \bm e_i(t)
\end{align}
of constant magnitude $F^a$. We consider this as an external force in contrast to the self-propulsion force of Ref.~\cite{mart:19}. As a consequence, an individual monomer in a fluid is no longer active force free, but the latter  gives rise to a Stokeslet \cite{elge:15}. In any case---for polymers with self-propelled or externally-driven  monomers---, Stokeslets appear by the forces of the potentials \eqref{eq:pot_bond}-\eqref{eq:pot_lj} and thermal noise.  
As for an active Brownian particle, we set $F^a= \gamma v_0$, with the friction coefficient $\gamma= 3\pi \eta d_H$ of the surrounding fluid---$\eta$ is the fluid viscosity and $d_H$ the monomer hydrodynamic diameter---and the active velocity $v_0$.
The orientation $\bm e_i$ changes in a diffusive manner according to
\begin{align} \label{eq:orientation}
\dot{\bm{e} }_i(t)= & \ \hat{\bm{\eta}}_i(t)\times  \bm{e}_i(t) ,
\end{align}
where $\hat{\bm \eta}_i$ is a Gaussian and Markovian stochastic processes with zero mean and the second moments
\begin{align}
\lla \hat \eta_{i \alpha}(t) \hat \eta_{j \beta}(t') \rra & = 2 D_R \delta_{\alpha \beta} \delta_{ij} \delta(t-t') .
\end{align}
Here, $T$ denotes the temperature, $k_B$ the Boltzmann constant, $D_R$ the rotational diffusion coefficient of a spherical colloid of diameter $d_H$, and $\alpha, \beta \in \{x,y,z \}$ refer to the axis of the Cartesian reference frame.

Fluid mediated interactions are incorporated implicitly  by  the Rotne-Prager-Yamakawa (RPY) hydrodynamic tensor \cite{rotn:69,yama:70} or explicitly by modeling the fluid via the MPC approach \cite{kapr:08,gomp:09}.

\subsection{Brownian dynamics with the RPY tensor} \label{sub_sec:model_BD}

In Brownian dynamics simulations in presence of hydrodynamic interactions, the overdamped equations of motion
\begin{align} \label{eq:langevin_bd}
\dot {\bm r}_i(t) = \sum_{j=1}^{N_m} \mathrm{\bf H}_{ij} \left[ \bm F^a_j + \bm F_j + \bm \varGamma_j (t) \right]
\end{align}
are considered. The forces $\bm F_i = - \bm \nabla_{\bm r_i} (U_l + U_b +U_{LJ})$ follow from the potentials \eqref{eq:pot_bond}-\eqref{eq:pot_lj}, and
$\bm \varGamma_i$ accounts for thermal fluctuations. The random force $\bm \varGamma_i$ is modeled as a Gaussian and Markovian stochastic processes with zero mean and the second moments
\begin{align}
\lla \bm \varGamma_i (t) \bm \varGamma_j^T (t')\rra & = 2 k_BT \mathbf{H}_{ij}^{-1} \delta(t-t') \ ,
\end{align}
where $\bm \varGamma_i^T$ denotes the transpose of $\bm \varGamma_i$ and $\mathbf{H}_{ij}^{-1}$ the inverse of $\mathbf{H}_{ij}$. The hydrodynamic tensor, $\mathbf{H}_{ij}(\bm r_{ij})$, is given by
\begin{align}
\mathbf{H}_{ij}(\bm r_{ij}) = \frac{\delta_{ij}}{3 \pi \eta d_H} \mathbf{I} + (1-\delta_{ij}) \bm{\Omega}(\bm r_{ij}) ,
\end{align}
where the first term on the right-hand side accounts for local friction and the RPY tensor $\bm{\Omega}(\bm r_{ij})$ for inter-particle interactions \cite{doi:86,mart:19}.
The RPY tensor ensures the positive definiteness of the hydrodynamic tensor even at small distances.
The translational equations of motion \eqref{eq:langevin_bd} are solved via the Ermak-McCammon algorithm \cite{erma:78}. The procedure to solve the equations of motion (\ref{eq:orientation}) for the orientation vector is described in Ref. \cite{wink:15}.

The active noise is quantified by the dimensionless P\'eclet number \cite{eise:16,das:18.1}
\begin{align} \label{eq:peclet}
 Pe = \frac{v_0}{l D_R},
\end{align}
which compares the time for the reorientation of an ABP monomer with that for its translation with velocity $v_0$ over the monomer radius.
The ratio between translational, $D_T=k_BT/3\pi \eta d_H$,  and rotational diffusion, $D_R$, of a single monomer is denoted as
\begin{align} \label{eq:delta}
  \Delta = \frac{D_T}{d_H^2 D_R} .
\end{align}
In the follwing, we will always consider $\Delta = 0.6$. 
The coefficient  $\kappa_{l}$ (Eq.~\eqref{eq:pot_bond}) for the bond strength is adjusted according to the applied P\'eclet number, in order to avoid bond stretching with increasing activity. By choosing $\kappa_{l} l^2 /k_BT = (10  +  2 Pe ) 10^3$,  bond-length variations are smaller than $3 \%$ of the equilibrium value $l$. Furthermore, the scaled bending force coefficient $\tilde{\kappa}_{b}=\kappa_{b} l^2 /k_BT$  (Eq.~\eqref{eq:pot_bend}) is related to the polymer persistence length, $l_p=1/(2p)$, by
\begin{equation}
pL = N_m \frac{\tilde{\kappa}_{b} \left( 1 - \coth \left( \tilde{\kappa}_{b} \right) \right) +1}{
	\tilde{\kappa}_{b} \left( 1 + \coth \left( \tilde{\kappa}_{b} \right) \right) -1} \ \  .
\end{equation}
The parameters of the truncated and shifted Lennard-Jones potential are $\sigma = 0.8 l$ and $\epsilon = k_BT$.

\subsection{Active polymers in  MPC fluid}\label{sub_sec:model_mpc}

\subsubsection{Polymer dynamics}

Every monomer is exposed to an active forces $\bm F_i^a=v_0\bm e_i(t)$ \eqref{eq:active_force}, hence, a polymer experiences the total external force
\begin{align}\label{eq:active-force-mpc}
\bm F^a  = \sum_{i=1}^{N_m} \gamma  v_0\bm e_i(t) = \frac{Pe}{d_H\Delta} \sum^{N_m}_{i=1}\bm e_i(t) ,
\end{align}
which drags along fluid and induces an overall fluid flow \cite{sing:18}. In a confined systems, \rev{walls prevent  global flow and give rise to fluid backflow.} To prevent a net fluid flow in our system with periodic boundary conditions, we modify the equations of motion of the fluid (and the embedded polymer) in such a way that the total momentum of the system (fluid plus polymer) vanishes \cite{sing:18}. This implies the backflow force on a monomer:
\begin{align} \label{eq:backflow-force-mpc}
\bm F^{b}_i  = -\frac{M}{mN +MN_m}\bm F^{a},
\end{align}
where $m$ is the mass of the fluid particle,  $N$ is the total number of fluid particles, and $M$ is the mass of a monomer.
The dynamics of a monomer is then described by the equation of motion
\begin{align} \label{eq:newton}
M \ddot{\bm r}_i = \bm F_i + \bm F_i^a + \bm F_i^{b} ,
\end{align}
with the force $\bm F_i$ following from the potentials \eqref{eq:pot_bond}-\eqref{eq:pot_lj}. Equation~\eqref{eq:newton} is solved by applying the velocity-Verlet algorithm.

\subsubsection{Fluid dynamics and fluid-polymer coupling}

The dynamics of the MPC fluid proceeds in two steps---streaming and collision \cite{kapr:08,gomp:09}. In the steaming step, Newton's equations of  motion  for  fluid particles are solved in the presence of the backflow force
$m \bm F_i^{b}/M$ over a time interval $h$, denote as collision time. Since $\bm e_i(t)$ changes very slowly in the time interval $h$ for small diffusion coefficients $D_R$, we apply the integration scheme
\begin{align} \label{eq:active-stream-mpc}
\bm v_k(t+h) & = \bm v_k(t)  - \frac{h}{mN +MN_m} \bm F^a(t) , \\
\bm r_k(t+h) & =\bm r_k(t)  + h\bm v_k(t)  - \frac{h^2}{2(mN +MN_m)} \bm F^a(t)
\end{align}
where $\bm r_k(t)$ and $\bm v_k(t)$ are the position and velocity of the MPC particle $k$ at time $t$, respectively.
In the collision step, particles are sorted into cubic cells of side length $a$ of a cubic, periodic systems of volume  $V=Na^3/\langle N_c \rangle$ to define the collision environment; $\langle N_c \rangle$ is the mean number of fluid particles in a collision cell. Subsequently, the relative velocity of each particle, with respect to the center-of-mass velocity of all the particles within the corresponding collision cell, is rotated by a constant angle $\alpha$ around a arbitrarily orientated axis. The orientation of the rotation axis is chosen randomly and independently for every cell and collision step. Hence, the final velocity after a MPC step is
\begin{align} \label{eq:collision-mpc}
\bm v_k(t) = \bm v_{cm}(t) + \mathrm{\bf R} (\alpha) \left[ \bm v_k(t) - \bm v_{cm}(t)  \right] ,
\end{align}
where $\mathrm{\bf R} (\alpha)$ is the rotation matrix, and
\begin{align}
\bm v_{cm}(t) = \frac{\sum^{N_c}_{k=1} m \bm v_k(t) + \sum_{j=1}^{N_m^c} M \bm v_j(t)}{m N_c + M N_m^c}
\end{align}
is the center-of-mass velocity of the $N_c$ MPC particles and the $N_m^c$ monomers within the cell of particle $k$. Similarly to Eq.~\eqref{eq:collision-mpc}, the velocities of the monomers are rotated, which yields the fluid-monomer coupling by MPC collisions.

Partitioning of space in collision cells implies violation of Galilean invariance, which is reinstalled by a random shift of the collision lattice at every collision step \cite{gomp:09,ihle:03}. In order to maintain locally a constant temperature, the Maxwell-Boltzmann scaling method is applied \cite{huan:10.1}. \\

We measure energies in units of $k_BT$, lengths in units of the collision cell  $a = l$, which is set equal to the equilibrium bond length, and time in units of $\tau=\sqrt{ma^2/k_BT}$. The  MPC particle mass is set to $m=1$, the monomer mass to $M=10m$,  the average number of particles in a  collision cell to $\langle N_c \rangle = 10$, and $\epsilon = k_BT$. A time step $h=0.01\sqrt{ma^2/k_BT}$ is used, which corresponds to the viscosity $\eta = 82.14 \sqrt{m k_BT/a^4}$ \cite{wink:09}. 
MPC is an ideal gas and, hence, its isothermal velocity of sound is $c_T = \sqrt{k_BT/m}$, which is unity in the units of the simulation. To realize low Mach numbers, the transport velocity of an active monomer has to be small compared to $c_T$. All simulations are performed in a cubic periodic box of linear size $L_B=100a$.

In order to compare simulation results obtained via the MPC approach with the Brownian dynamics simulations using  the RPY tensor, several parameters have to be adjusted. In particular, MPC simulations yield the hydrodynamic diameter $d_H =0.6a$ of a monomer \cite{pobl:14,sing:14}, which yields, with $D_R=100/\tau$, $\Delta = k_BT/(3 \pi \eta d_H^3 D_R) \approx 0.6$.

\section{Computer simulations: Results} \label{sec:results_sim}

\subsection{Conformational properties}

\begin{figure}[t]
\begin{center}
     \includegraphics[width=\columnwidth]{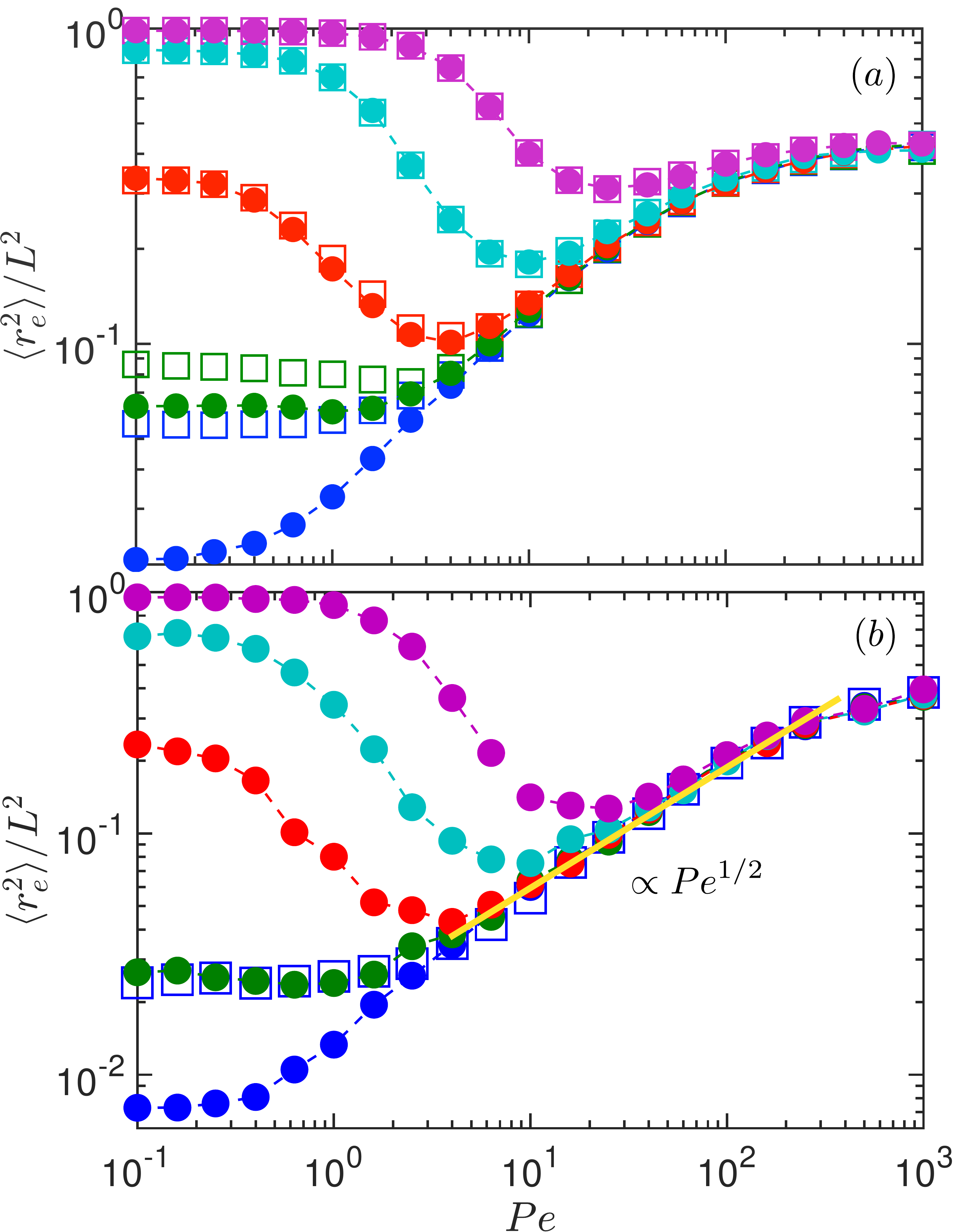} 
\end{center}
\caption{Polymer mean square end-to-end distance as a function of the P\'eclet number for semiflexible polymers with (a)  $N_m=50$ ($L=49l$)  and (b) $N_m=150$ ($L=149l$)   monomers.  Bullets are results of phantom polymers and squares results of self-avoiding polymers in (a)  for $pL = 5 \times 10^1$ (blue), $1.5 \times 10^1$ (green), $2.6$ (red), $2.5 \times 10^{-1}$  (cyan), and $2.5 \times 10^{-2}$ (purple), and in (b) for $pL = 1.5 \times 10^2$ (blue), $4.5 \times 10^1$ (green), $7.5$ (red), $7.5 \times 10^{-1}$  (cyan), and $7.5 \times 10^{-2}$ (purple) (bottom to top). The dashed lines are guides for the eye. The solid line (yellow) in (b) indicates a power-law dependence in the respective  regime.
Hydrodynamics is taken into account by the RPY hydrodynamic tensor.
} \label{fig:end_ED_BD}
\end{figure}

The average shape of the polymers are characterized by their mean square end-to-end distance. Figure~\ref{fig:end_ED_BD} displays results for phantom and self-avoiding polymers of lengths $L=(N_m - 1)l = 49l, \ 149l$ and various persistence lengths $l_p=1/(2p)$. Evidently, flexible polymers, with $pL \gg 1$, swell monotonically with increasing P\'eclet number, whereas semiflexible polymers shrink at moderate $\mathrm{Pe}$, and swell for large $\mathrm{Pe}$ similarly as flexible polymers. In the  asymptotic limit  $\mathrm{Pe} \rightarrow \infty$, the value $\langle \bm r_e^2 \rangle  \approx 2L^2/5$ is assumed. Excluded-volume interactions change the behavior in so far as $\langle \bm r_e^2 \rangle/L^2$ starts at a larger equilibrium value (cf. Fig.~\ref{fig:end_ED_BD}(a)). For higher $\mathrm{Pe}$ and swollen polymers, self-avoidance becomes irrelevant. A qualitative similar behavior is obtained for longer polymers, only quantitative differences  appear (cf. Fig.~\ref{fig:end_ED_BD}(b)). However, longer polymers exhibit the universal, persistence-length independent increase  $\langle \bm r_e^2 \rangle \sim \mathrm{Pe}^{1/2}$ with increasing $\mathrm{Pe}$ above a critical value. This  regime appears for sufficiently long polymers only and is not present for $N_m=50$.  In addition, a non-universal cross-over regime exists for flexible polymers in the range $5 \lesssim\mathrm{Pe} \lesssim 30$. These regimes and the observed $\mathrm{Pe}$-dependence of the universal regime are explained by the theoretical model in Sec.~\ref{sec:conformations}. 

\begin{figure}[t]
	\begin{center}
		\includegraphics[width=\columnwidth]{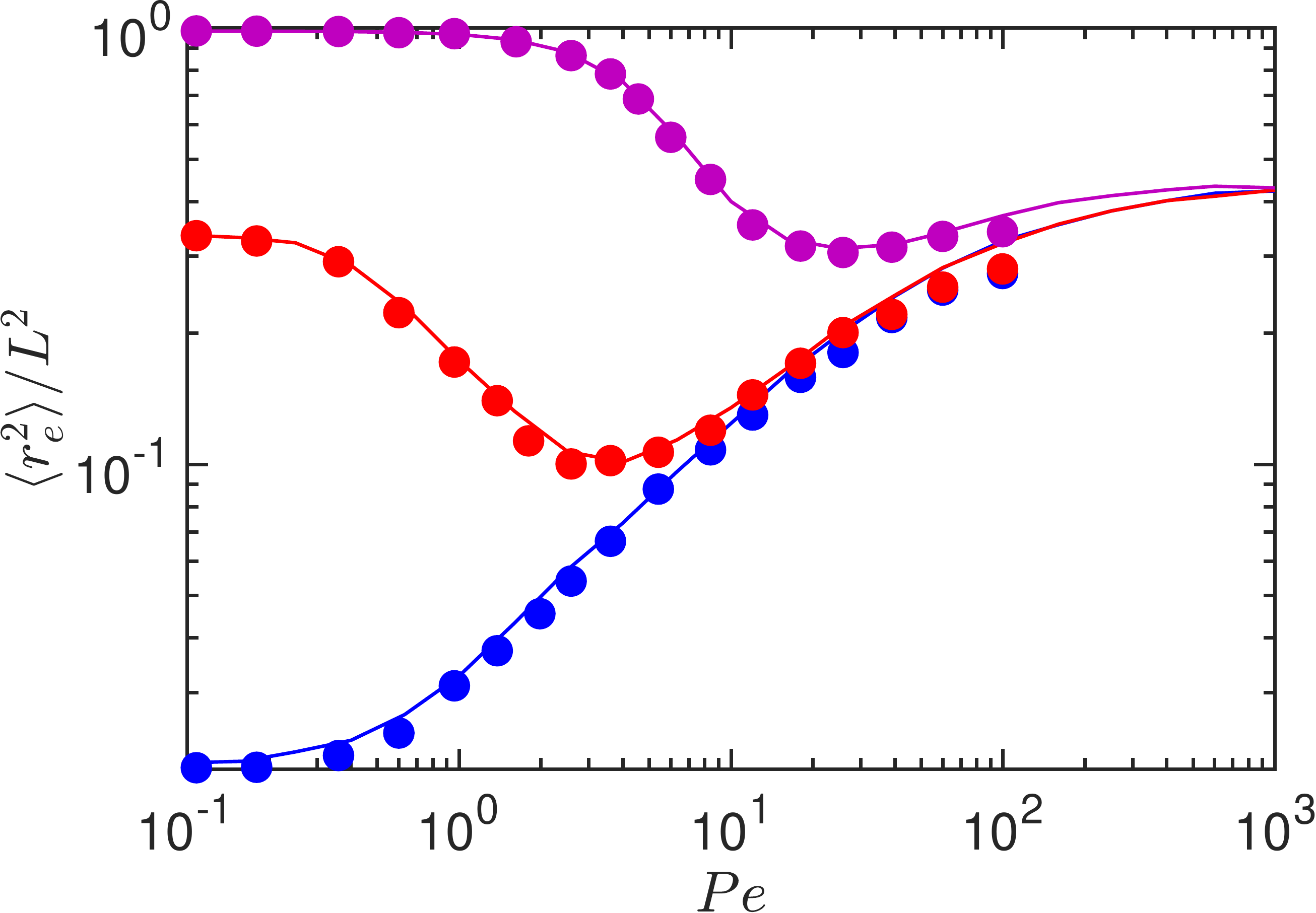}
	\end{center}
\caption{Polymer mean square end-to-end distance as a function of the P\'eclet number of semiflexible polymers with $N_m=50$ ($L=49l$) monomers for  $pL = 5 \times 10^1$ (blue),  $2.6$ (red), and $2.5 \times 10^{-2}$ (purple) (bottom to top). Solid lines are results applying the RPY tensor and bullets are results of hydride simulations using the MPC approach.} \label{fig:end_ED_MPC_BD}
\end{figure}

The mean square end-to-end distances obtained for polymers embedded in a MPC solvent are compared with the simulation results applying the RPY tensor in Fig. \ref{fig:end_ED_MPC_BD}. Good quantitative agreement of the polymer conformations for the two simulation approaches is obtained, which confirms their suitability for these simulation studies. For the hybrid MPC approach, deviations from the RPY tensor simulations appear for $\mathrm{Pe} \gtrsim 10^2$. This is attributed to limitations of the MPC approach in terms of Mach and Reynolds numbers. The range of P\'eclet number can be extended by applying a smaller collision time step and/or by a higher mean value of MPC particles in a collision cell. 

\begin{figure}[t]
\begin{center}
	\includegraphics[width=\columnwidth]{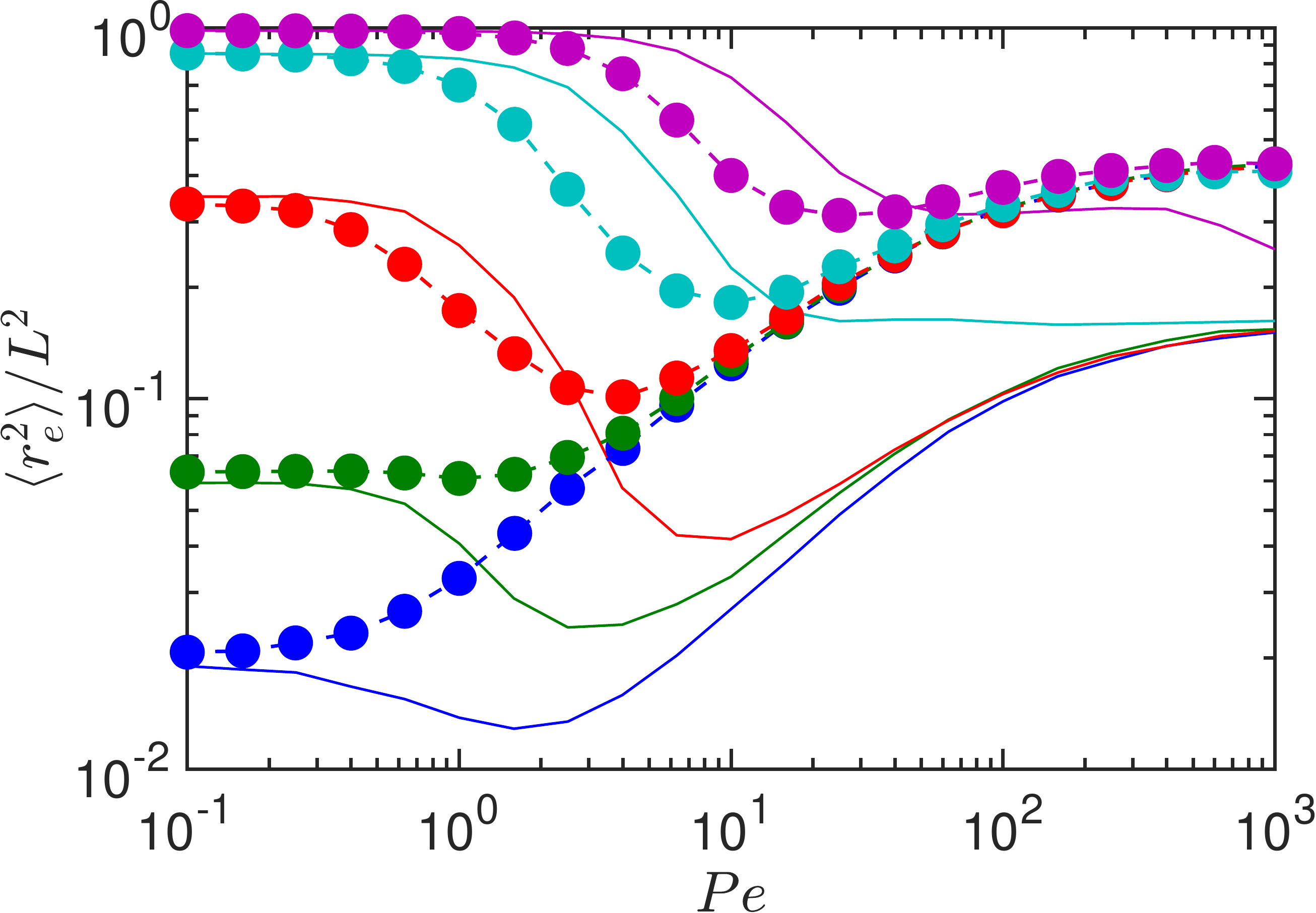}
\end{center}
\caption{Polymer mean square end-to-end distance as a function of the P\'eclet number of semiflexible polymers of length  $N_m=50$ ($L=49l$)  and $pL = 5 \times 10^1$ (blue), $1.5 \times 10^1$ (green), $2.6$ (red), $2.5 \times 10^{-1}$  (cyan), and $2.5 \times 10^{-2}$ (purple) (bottom to top).  Bullets indicate results for the external active process and solid lines the respective results for self-propelled monomers  (ABPO+HI \cite{mart:19}). Hydrodynamics is taken into account by the RPY hydrodynamic tensor.} \label{fig:end_external_self}
\end{figure}

The structural properties of the polymer in presence of hydrodynamic interactions strongly depends on the nature of the active process. As discussed in Sec.~\ref{sec:active_polymer},   the active force is considered here  as an external force, mimicking an active environment. Figure~\ref{fig:end_external_self} shows that such an external active force leads to a significantly stronger polymer swelling than intrinsic self-propulsion (cf.  Ref.~\cite{mart:19}). Remarkably, in contrast to  the shrinkage of flexible active polymers with self-propelled monomers  over a range of P\'eclet numbers, flexible externally-driven active polymers  monotonically  swell. Moreover, the externally-driven active polymers assume a larger asymptotic mean square end-to-end distance for $\mathrm{Pe} \to \infty$, i.e., intrinsically active Brownian polymers in presence of hydrodynamic interactions are  more compact. We will provide a qualitative and quantitative explanation for these observations in Sec.~\ref{sec:conformations}.

\subsection{Dynamical properties}

\begin{figure}[t]
	\begin{center}
		\includegraphics[width=\columnwidth]{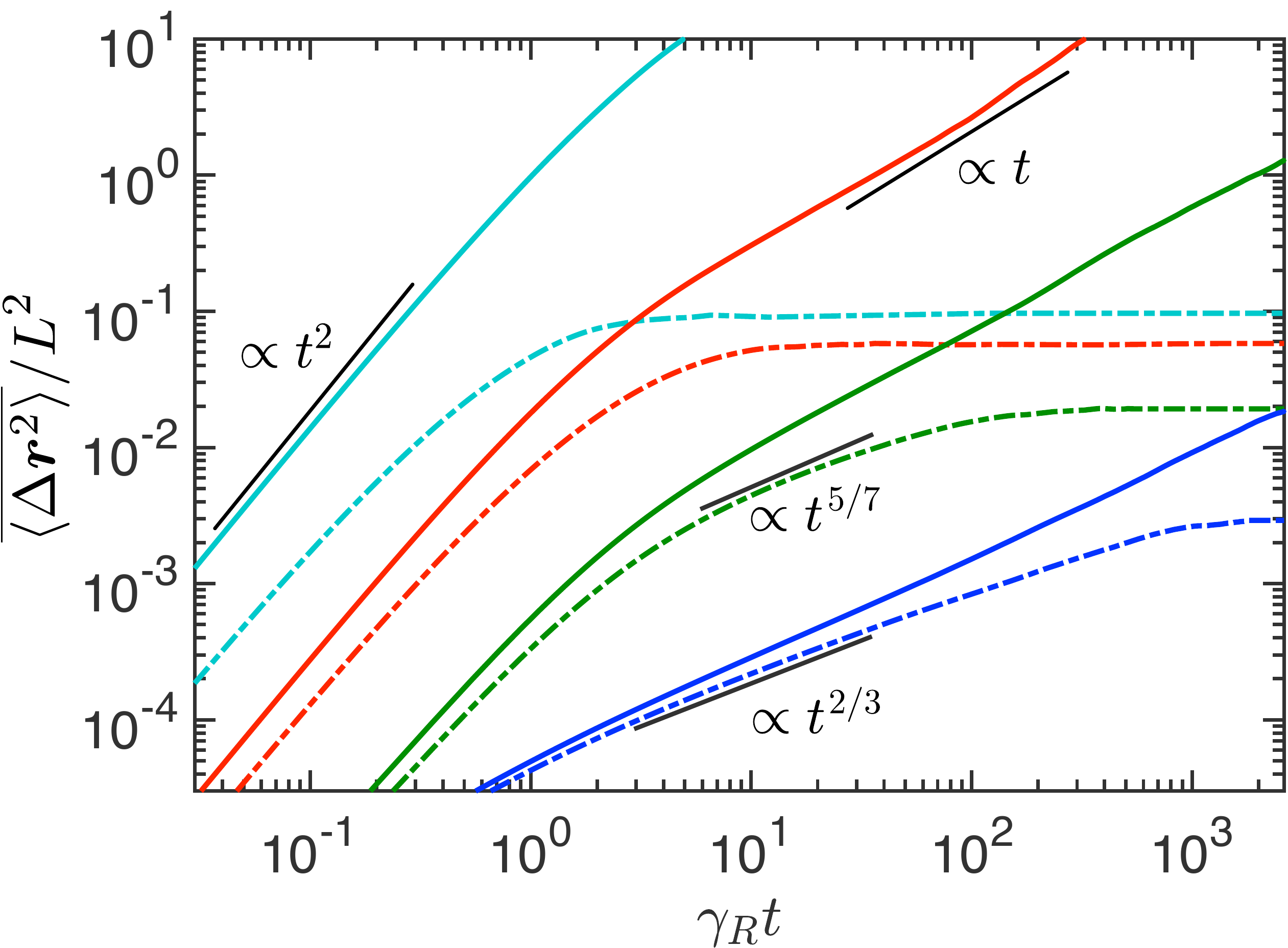}		
    \end{center}
\caption{Mean square displacement of flexible polymers with $N_m= 150$ ($pL = 150$) monomers for the P\'eclet numbers $\mathrm{Pe} = 0$ (blue), $10^1$ (green), $10^2$ (red), and $10^3$ (cyan) (bottom to top). The time is scaled by the factor $\gamma_R = 2D_R$. The solid lines indicate the overall monomer MSD and the dashed lines their  MSD in the polymer center-of-mass reference frame. The short  lines (black)  indicating  a  power-law dependence of the data in the respective regime. \rev{Hydrodynamics is taken into account by the RPY hydrodynamic tensor.}} \label{fig:msd_sim}
\end{figure}

The effect of activity on the polymer dynamics is illustrated in Fig.~\ref{fig:msd_sim}, which displays the average monomer mean square displacement
\begin{align}  \label{eq:def_msd_dis}
 \lbar {\lla  \Delta \bm r^2(t) \rra} =  \ \frac{1}{N_m} \sum_{i=1}^{N_m} \lla  (\bm r_i(t) - \bm r_i(0))^2 \rra .
\end{align}
A passive polymer exhibits the well-known Zimm behavior for $t/\tilde \tau_1 \ll   1$, with the time dependence $t^{2/3}$ of the MSD in the center-of-mass reference frame, where $\tilde \tau_1$ is the longest polymer relaxation time in presence of HI \cite{doi:86}. In the asymptotic limit $t \to \infty$, the MSD depends linearly on time, with an activity-dependent diffusion coefficient (cf. Sec.~\ref{sec:cont_dyn}). At large activities, the MSD exhibits a ballistic regime for short times, similar to  a single ABP \cite{elge:15}, but  with the reduced (average) velocity $v_0/\sqrt{N_m}$ (cf. Eq.~\eqref{eq:effective_diff_coef}). Due to the independence of the monomer rotational motion, the effective center-of-mass ballistic velocity is determined by the  fluctuations of the monomer propulsion direction, which yields the factor $1/\sqrt{N_m}$.
 In the center-of-mass reference frame, the monomer MSD exhibits a subdiffusive power-law regime for $\mathrm{Pe} \approx 10$, with an activity-determined effective exponent of $5/7$ (cf. Sec.~\ref{sec:cont_dyn} for a derivation of the exponent). This regime extents with polymer length, but becomes smaller with increasing $\mathrm{Pe}$, since the relaxation time $\tilde \tau_1$ decreases with increasing P\'eclet number. Nevertheless, it is a consequence of hydrodynamic interactions and activity (cf. Sec.~\ref{sec:cont_dyn} for a more detailed discussion).

\section{Analytical Approach} \label{sec:model-analy}

Insight into the qualitative differences between externally-driven active polymers and a polymers composed of self-propelled monomers is achieved by an analytical model, where the polymers are described as   continuous Gaussian semiflexible chains. This model has previously been applied to linear and ring active Brownian free-draining polymers \cite{eise:16,eise:17,eise:17.1,mart:18.1,mous:19}, as well as to linear self-propelled polymers with hydrodynamic interactions \cite{mart:19}.

\subsection{Model and equations of motion}

The polymers are  considered as differentiable space curves $\bm r (s,t)$ of total length $L$, with contour coordinate $s \left( -L/2 \leq s \leq L/2 \right)$, and their conformations change with time $t$. The external active process is introduced by assigning an independent velocity $\bm v (s,t)$ to every site $\bm r (s,t)$. The corresponding Langevin equation  is \cite{harn:96,petr:06}
\begin{align} \label{eq:langevin_analy}
& \frac{\partial \bm r(s,t)}{\partial t}  =   \int_{-L/2}^{L/2} \! ds' \,  \brh (\bm r(s),\bm r(s'))  \bigg[ 3 \pi \eta \bm v (s',t)     \\ \nonumber &  \hspace*{1cm}+   2\nu k_BT \frac{\partial^2 \bm r (s',t )}{\partial s'^2} - \epsilon k_BT \frac{\partial^4 \bm r (s',t)}{\partial s'^4} + \bm \varGamma(s',t) \bigg] .
\end{align}
Free-end boundary conditions are applied as described in Refs.~\cite{harn:95,eise:16}. Moreover, the constraint on the (average) contour length
\begin{align} \label{eq:constraint_general}
\int_{-L/2}^{L/2} ds \lla \left( \frac{\partial \bm r(s,t)}{\partial s} \right)^2 \rra  = L
\end{align}
is take into account, which is fundamental to achieve the correct polymer properties \cite{harn:95,eise:16,mous:19}.
The tensor 
\begin{align}  \label{eq:hydro_tensor}
{\bf H}({\bm r}(s),{\bm r}(s')) =  \frac{\delta(s-s')}{3 \pi \eta}{\bf I}  +  {\bm \Omega}({\bm r}(s)-{\bm r}(s')) 
\end{align}
captures the hydrodynamic interactions, where the first term on the right-hand side describes the local friction, and
\begin{align} \label{eq:oseen_tensor}
{\bm \Omega}(\Delta \bm r) = \frac{1}{8 \pi \eta | \Delta \bm r |} \left( {\bf I} + \frac{\Delta \bm r  \otimes \Delta \bm r }{|\Delta \bm r|^2} \right) \,
\end{align}
is the Oseen tensor \cite{doi:86,mart:19}. The terms in Eq.~(\ref{eq:langevin_analy}) with the second and forth derivative capture chain flexibility, i.e., chain entropy, and bending forces, respectively.  The Lagrangian multiplier $\nu$ accounts for the inextensibility of the polymer (we will denote $\nu$ as stretching coefficient in the following) and $\epsilon$ characterizes the bending stiffness \cite{wink:94,wink:03}. For a polymer in three dimensions, previous studies yield $\epsilon =
3/4p$ \cite{wink:94,wink:03}.

For the velocity $\bm v(s,t)$, we adopt a non-Markovian Gaussian stochastic processes with zero mean and the correlation function (colored noise)
\begin{align} \label{eq:corr_colored}
\lla \bm v(s,t) \cdot \bm v(s',t')\rra = v_0^2 l e^{-\gamma_R |t-t'|} \delta(s-s') \ .
\end{align}
This  correlation function follows from Eq.~\eqref{eq:orientation} or, similarly, by considering a monomer as an active Ornstein-Uhlenbeck particle (AOUP) \cite{elge:15,wink:16,eise:16,sama:16,das:18.1}.

As outlined in Sec.~\ref{sec:active_polymer},  $\bm v (s,t)$ is a consequence of an external forcing, hence $\bm v(s,t)$ appears inside the integral in Eq.~\eqref{eq:langevin_analy} and implies a Stokeslet flow. 

\subsection{Solution of the equations of motion}

\subsubsection{Hydrodynamic tensor: Preaveraging approximation}

In order to find an approximate analytical solution of the nonlinear and nonlocal equation of motion \eqref{eq:langevin_analy}, we apply the preaveraging approximation, where the hydrodynamic tensor $\brh (\bm r(s)-\bm r(s'))$ is replaced by its stationary-state average, i.e., ${\bf H}(\bm r(s) - \bm r(s')) \rightarrow \lla{\bf H}(\bm r(s) - \bm r(s')) \rra = {\bf H}(s,s')$ \cite{doi:86,harn:96}. Hence, Eq.~(\ref{eq:langevin_analy}) turns into a linear equation (Ornstein-Uhlenbeck process) with a Gaussian stationary-state distribution function for the distance $\Delta \bm r(s,s') = \bm r(s) - \bm r(s')$ of the form \cite{das:18.1,harn:96,doi:86}
\begin{align} \label{eq:stat_state}
\Psi( \Delta \bm r ) = \left(\frac{3}{2\pi a^2(s,s') }\right)^{3/2}\exp \left(-\frac{3\Delta \bm r^2}{2 a^2(s,s')} \right) \, ,
\end{align}
with $a^2(s,s') = \lla (\bm r(s) - \bm r(s'))^2 \rra$. Then, the Oseen tensor \eqref{eq:oseen_tensor} becomes
\begin{align}\label{eq:oseen_pre_avr}
{\bm \Omega}(s,s')
= \frac{\Theta (|s-s'|-d_H)}{3 \pi \eta} \sqrt{\frac{3}{2 \pi a^2}} {\bf I} = \Omega (s,s') {\bf I}.
\end{align}
The Heaviside step function $\Theta(x)$ introduces $d_H$ as a lower cut-off for the hydrodynamic interactions, which can be identified with the thickness of the polymer.

The preaveraging approximation has very successfully been applied to describe the dynamics of DNA \cite{petr:06} and semiflexible polymers \cite{harn:96}. Even quantitative agreement between analytical theory and simulations of the full hydrodynamic contribution of rather stiff polymers is achieved \cite{hinc:09}, as well as with measurements on DNA \cite{petr:06}. This demonstrates the suitability of preaveraging even for stretched polymers. However, the preaveraging approximation overestimates the hydrodynamics of rodlike objects \cite{wink:07.1}.

\subsubsection{Eigenfunction expansion}

The linearized equation of motion is solved by the eigenfunction expansion
\begin{align} \label{eq:eigen_expand}
\bm r(s,t) = \sum_{n=0}^{\infty} \bm \chi_n (t) \vphi_n(s),
\end{align}
in terms of the eigenfunctions $\vphi_n$ of the equation
\begin{align}
\epsilon k_BT \frac{d^4}{ds^4} \vphi_n(s) - 2\nu k_BT \frac{d^2}{ds^2} \vphi_n(s) = \xi_n \vphi_n(s) \, ,
\end{align}
with the eigenvalues ($n \in \mathbb{N}_{0}$)
\begin{align} \label{eq:eigenvalue}
\xi_n = k_BT \left(\epsilon \zeta_n^4+ 2\nu \zeta_n^2 \right) .
\end{align}
The wave numbers $\zeta_n$ follow from the boundary conditions. For a passive flexible polymer, $pL \gg 1$, the wave numbers are $\zeta_n = n \pi/L$ and the eigenvalues $\xi_n = 2 \nu k_BT \pi^2 n^2/L^2$.  The stiffness dependence of $\zeta_n$ and $\xi_n$ of passive semiflexible polymers is discussed in Ref.~\cite{harn:95} and for free draining active polymers in Ref.~\cite{eise:17}.

Insertion of the expansion \eqref{eq:eigen_expand} into  Eq.~\eqref{eq:langevin_analy} yields the equation
\begin{align} \label{eq:chi_eq}
\frac{d  {\bm \chi}_n (t)}{d t } = \sum_{m=0}^{\infty}  H_{nm}\left[\gamma\bm v_m(t) + \bm \varGamma_m (t) - \xi_m  {\bm \chi}_m (t) \right]
\end{align}
for the mode amplitudes $\bm \chi_n$,
where $H_{nm} = (\delta_{nm} + 3 \pi \eta \Omega_{nm} )/3\pi \eta$ is the hydrodynamic tensor in  mode representation \cite{harn:96}.
The second moments of the stochastic-force amplitudes $\bm \varGamma_n(t)$ are given by
\begin{align} \label{eq:corr_stoch_mode}
\langle  \varGamma_{n \alpha}(t) \, \varGamma_{m \beta}(t')  \rangle & = \, 2 k_B T \delta_{\alpha \beta } \delta(t-t') \, H_{nm}^{-1} \, .
\end{align}
The mode representation of the correlation function (\ref{eq:corr_colored}) of the active velocity is \cite{eise:16}
\begin{align} \label{eq:mode_corr_v}
\langle \bm v_n(t) \cdot \bm v_m (t') \rangle = \, v_0^2 l e^{-\gamma_R |t-t'|} \delta_{nm} \, .
\end{align}

In Eq.~(\ref{eq:chi_eq}), all modes couple in general and the set of equations can only be solved numerically. To arrive at an analytical solution, we neglect the off-diagonal terms of the hydrodynamic-mode tensor $H_{nm}$, which yields \cite{doi:86,harn:96,petr:06} ($n>0$)
\begin{equation} \label{eq:chi_eqMotHydUncoup}
\frac{d \bm \chi_n (t)}{d t } = - \frac{1}{\tilde \tau_n} \bm \chi_n + H_{nn} \left [ \bm \varGamma_n (t)  +  \gamma \bm v_n(t) \right] \, ,
\end{equation}
with the relaxation times
\begin{align} \label{eq:relax_time}
\tilde{\tau}_n = \frac{1}{H_{nn} \xi_n} = \frac{\tau_n}{1 + 3 \pi \eta \Omega_{nn}} ,
\end{align}
and $\tau_n=3 \pi \eta/\xi_n$  the relaxation times in absence of hydrodynamic interactions. For passive flexible polymers \cite{harn:95,eise:17}
\begin{align} \label{eq:relax_flex}
\tau_n= \frac{3 \eta L^2}{2 \nu k_BT \pi n^2} .
\end{align}

The stationary-state solution of Eq.~(\ref{eq:chi_eqMotHydUncoup})  for  $n>0$ is
\begin{align} \label{eq:chi_n}
 \bm \chi_n (t) =  \ H_{nn} \int_{-\infty }^{t} \! dt' \, &e^{-(t-t') / \tilde{\tau}_n }  \left[ \gamma \bm v_n(t') +  \bm \varGamma_n (t') \right],
\end{align}
and for $n=0$
\begin{align} \label{eq:chi_0}
\bm \chi_0 (t)  =  \bm \chi_0(0)  + \int_{0 }^{t} \! dt' \, H_{00}  \left[ 3 \pi \eta \bm v_{(0)}(t') +  \bm \varGamma_0 (t')  \right].
\end{align}

\subsubsection{Correlation Functions}

The correlation functions of the mode amplitudes are given by ($n > 0$)
\begin{align} \label{eq:corr_mod_n}
\lla  \bm \chi_n(t)  \cdot   \bm \chi_m (t') \rra =  & \  \delta_{nm} \left( \frac{ k_BT \tau_n}{\pi \eta } e^{-|t-t'|/\tilde{\tau}_n } \right. \\ \nonumber & \hspace*{-2cm}
\left. + \frac{v_0^2 l {\tau}_n ^2}{1-(\gamma_R \tilde{\tau}_n)^2} \left[e^{-\gamma_R |t-t'|} - \gamma_R \tilde{\tau}_n e^{-|t-t'|/ \tilde{\tau}_n} \right] \right) \, ,
\end{align}
and for $n=0$
\begin{align} \label{eq:corr_mod_0}
&  \lla  \bm \chi_0(t) \cdot  \bm \chi_0(t')\rra  =    \lla \bm \chi_0^2 (0) \rra  + 6 k_B T \, H_{00} \, t' \\ \nonumber  &
+  (3 \pi \eta H_{00})^2 \frac{v_0^2 l }{\gamma_R^2}\left[ 2 \gamma_R t' -1 -e^{\gamma_R (t'-t)} + e^{-\gamma_R t} + e^{-\gamma_R t'} \right] \, .
\end{align}
Inserting the eigenfunction expansion (\ref{eq:eigen_expand})  into the mean square distance $a^2(s,s')$, we obtain
\begin{align} \label{eq:msqd}
a^2(s,s') = \sum_{n=1}^{\infty} \lla \bm \chi_n^2\rra (\vphi_n(s) - \vphi_n(s'))^2 ,
\end{align}
with the stationary-state correlation functions (\ref{eq:corr_mod_n})
\begin{align} \label{eq:chi_st_st}
\lla \bm \chi_n^2 \rra = \frac{k_BT \tau_n}{\pi \eta} + \frac{v_0^2l  \tau_n^2}{1+\gamma_R \tilde \tau_n} ,
\end{align}
which depend on the hydrodynamic interactions via $\tilde \tau_n$. The active term with $v_0^2$ leads to enhanced fluctuations, which are more significant at small mode numbers \cite{mous:19},  and reflects the violation of the fluctuation-dissipation relation \cite{gnes:18}.  Notably, hydrodynamic interactions affect the dynamics as well as the stationary-state conformational properties of an active polymer, in contrast to passive systems, where conformational properties are independent of HI.

\subsubsection{Mean square distance and hydrodynamic tensor: Mode representation}

The exact analytical expression of the mean square distance $a^2(s,s')$ \eqref{eq:msqd} for the flexible active polymer can be calculated by (numerically) performing the sum in Eq.~(\ref{eq:msqd}), where, in general,  $a^2(s,s')$  depends on $s$ and $s'$. However, the relaxation times $\tilde \tau_n$ are required,  which depend via $a^2(s,s')$ on the Oseen tensor. Hence, the double integral
\begin{small}
\begin{align} \label{eq:oseen_mode_dist}
\Omega_{nn}  = \sqrt{\frac{1}{6 \pi^3 \eta^2}}  \int_{-L/2 }^{L/2} \int_{-L/2 }^{L/2}   \Theta (|s-s'|-d_H)  \frac{\varphi_n(s) \varphi_n(s')}{\sqrt{a^2(s,s') }} ds' ds
\end{align}
\end{small}
has to be evaluated together with Eq.~\eqref{eq:msqd}  in an iterative and self-consistent manner, which  constitutes a major computational challenge. 

\begin{figure}[h]
	\begin{center}
		\includegraphics[width=\columnwidth]{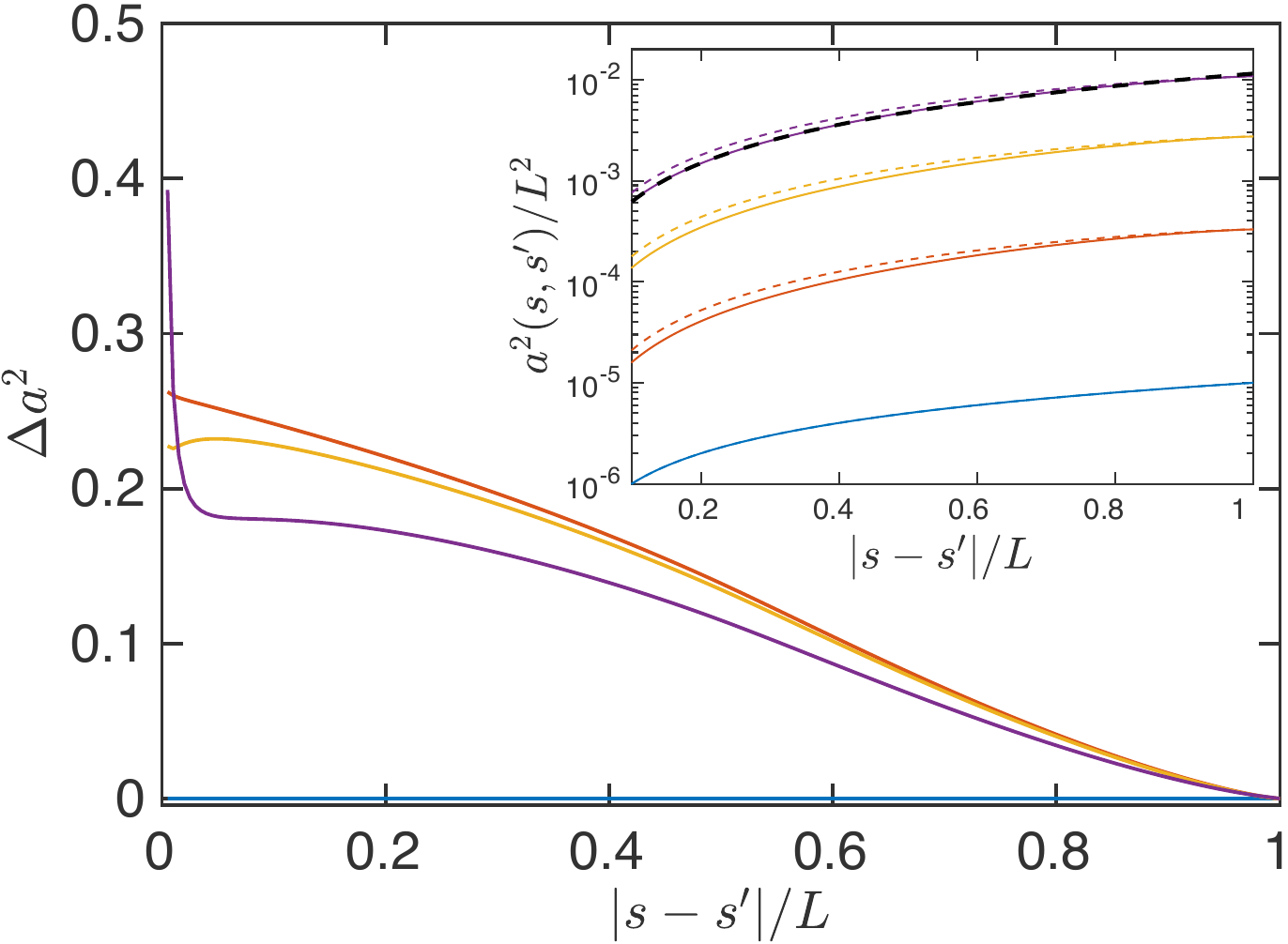}
	\end{center}
	\caption{ Ratio $\Delta a^2 =|a^2(s,s')- a^2(s-s')|/a^2(s,s')$ of the difference between the mean square distance between two points along the polymer contour,  Eq.~\eqref{eq:msqd}, and its approximation, Eq.~\eqref{eq:msqd_approx}, and Eq.~\eqref{eq:msqd} for  $pL=10^3$ and the P\'eclet numbers $\mathrm{Pe} =10^{-2}$ (blue, bottom), $1$ (orange), $50$ (yellow), and $10^3$ (purple) (top to bottom at 0.2). Inset:  Mean square distance between two points along the polymer contour. The solid lines are obtained from Eq.~\eqref{eq:msqd}, where $s' = -L/2$, and the dashed lines from the approximation  \eqref{eq:msqd_approx}. The long-dashed line for $\mathrm{Pe}= 10^3$ is a power-law fit, which yields $a^2(s) = 0.11 s^{1.27}$.   Colors correspond to the same P\'eclet numbers as in the main plot and increase from bottom to top.} \label{fig:sigmaHyd}
\end{figure}

For a passive semiflexible polymer,  $a^2(s,s')$ is only a function of the difference $|s-s'|$ \cite{wink:94,harn:96}. 
In order to find a  more easily tractable  expression for an active polymer,  we replace the difference of the eigenfunctions in Eq.~(\ref{eq:msqd}) by the expression valid for a passive polymer, namely  $\vphi_n(s)- \vphi_n(s')=2 \sin(n\pi(s-s')/2L)$ for $n$ odd  and $\vphi_n(s)- \vphi_n(s')=0$ for $n$ even.
As a result, we obtain the expression
\begin{align} \label{eq:msqd_approx}
a^2(s)= \frac{8}{L} \sum_{n, \text{odd} } \left( \frac{ k_BT \tau_n}{\pi \eta }    + \frac{v_0^2 l \tau_n ^2}{1+\gamma_R \tilde{\tau}_n} \right) \sin^2\left(\frac{n\pi }{2L} s \right)  \, .
\end{align}
This leads to the more easily tractable expression for the Oseen tensor \eqref{eq:oseen_mode_dist} with a single integral 
\begin{align} \label{eq:omega_single}
\Omega_{nn} =
\sqrt{\frac{2}{3\pi^3}} \frac{1}{\eta L}  \int_{d_H}^{L} \! \frac{L-s}{\sqrt{a^2(s)} } \cos \left( \frac{n\pi }{L} s \right) \, ds 
\end{align}
by applying a standard approximation for the double integral, which is dominated by contributions with $s=s'$ (cf. Ref.~\cite{doi:86}).
This expression is identical with that of a passive polymer aside from the distance $a^2(s-s')$, which depends here on activity via the relaxation times \cite{harn:96}.  As shown in Fig.~\ref{fig:sigmaHyd}, the approximations employed in deriving  Eq.~\eqref{eq:msqd_approx} capture the dependence of $a^2(s,s')$ on the contour coordinate well, the better the larger the P\'eclet number.

In the following, when not indicated otherwise, the approximate expressions \eqref{eq:msqd_approx} and \eqref{eq:omega_single} are used for the calculation of the Oseen tensor. Moreover, we use $\Delta = 0.6$ \rev{(cf. Eq.~\eqref{eq:delta} for the definition of $\Delta$)}.

\subsubsection{Stretching coefficient and relaxation times}

For flexible polymers with  $L/l=pL \gg 1$,  the constraint \eqref{eq:constraint_general} for the stretching coefficient $\mu = 2\nu /(3p)$  turns into
\begin{align} \label{eq:constraint}
\sum_{n=1}^{\infty} \left[ \frac{k_BT \tau_n}{\pi \eta} + \frac{v_0^2l \tau_n^2}{1+\gamma_R \tilde \tau_n} \right] \zeta_n^2 = L ,
\end{align}
with the eigenfunction expansion \eqref{eq:eigen_expand} and the relaxation times (Eq.~\eqref{eq:relax_time})
\begin{equation} \label{eq:tau_asym_flex}
\tilde{\tau}_n =  \frac{\tau_R}{\mu n^2 (1+3\pi \eta \Omega_{nn})} \ ,
\end{equation}
where $\tau_R = \eta L^2/(\pi k_B T p)$  is the  Rouse relaxation time \cite{harn:95,doi:86}. Due to nonlinear terms, specifically in $\Omega_{nn}$, the related equations and expressions have to be solved and evaluated numerically.

\begin{figure}[t!]
	\begin{center}
		\includegraphics[width=\columnwidth]{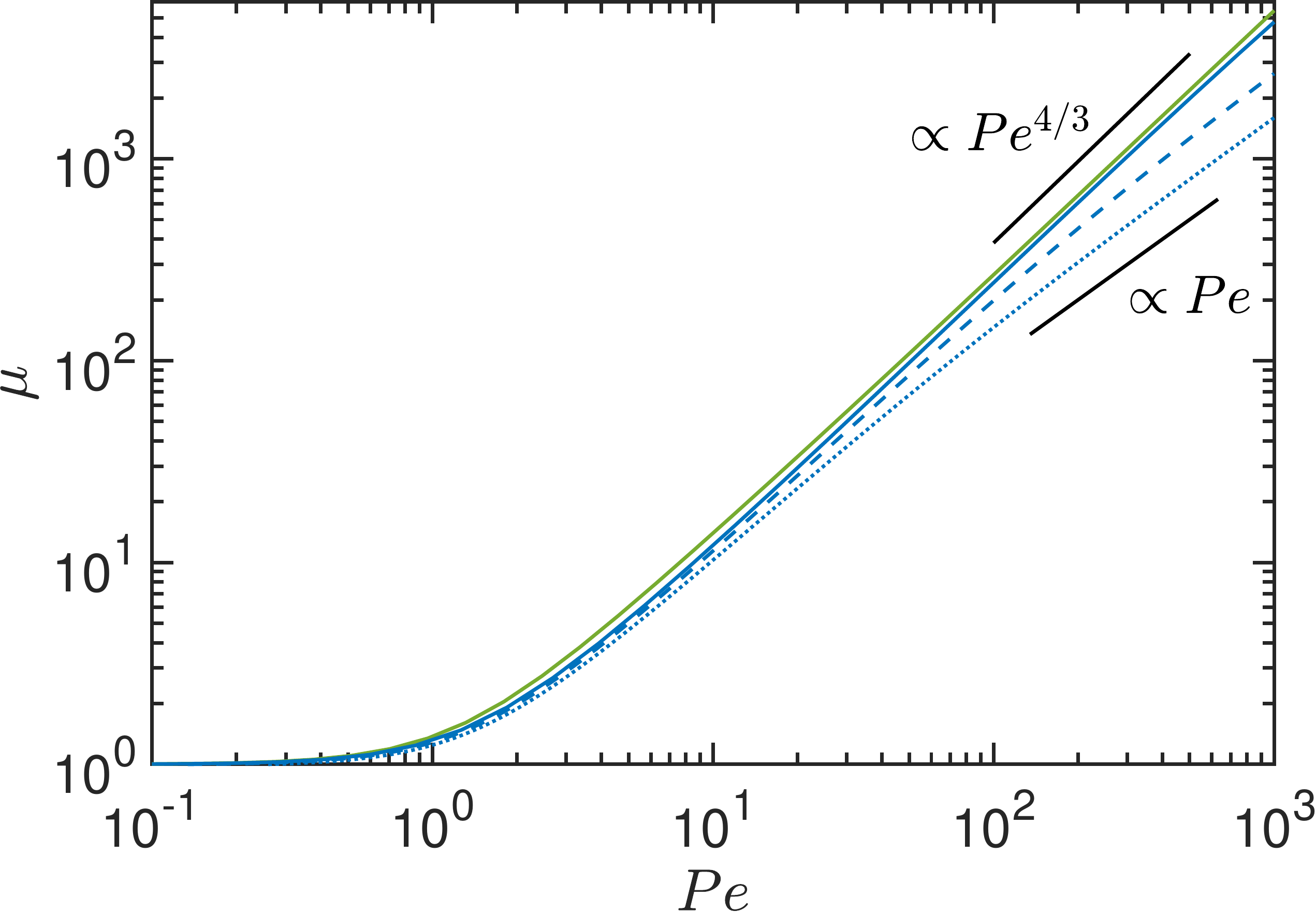}
	\end{center}
	\caption{Normalized stretching coefficient $\mu=2\nu /(3p)$, solution of Eq.~\eqref{eq:constraint}, as function of the P\'eclet number $\mathrm{Pe}$ for flexible polymers with $pL=50$ (dotted), $1.5 \times 10^2$ (dashed), and $10^3$ (solid blue, bottom). The top  solid line (green) shows the result of an active polymer in absence of HI for $pL=10^3$ (solid). The short lines (black) indicate the power-law dependence in the respective regimes.
	} \label{fig:lag}
\end{figure}

The scaled stretching coefficient, $\mu=2\nu /(3p)$, is presented in Fig.~\ref{fig:lag} as a function of the  P\'eclet number. For short polymers or larger stiffness ($pL=50$), $\mu$ increases linearly with increasing $\mathrm{Pe}$ in the limit $1 \ll pL \ll \mathrm{Pe}$ (cf. Eq.~\eqref{app:mu_linear}). In case of more flexible polymers ($pL\gtrsim 10^3$),  $\mu \sim \mathrm{Pe}^{4/3}$ in the range $1 \ll \mathrm{Pe} \ll pL$ (cf. Eq.~\eqref{app:mu_nonlinear}). The overall dependence of $\mu$ on $\mathrm{Pe}$ resembles that of a polymer in the absence of hydrodynamic interactions in this limit \cite{eise:16}. Yet, hydrodynamics affects $\mu$, particularly for P\'eclet numbers in the vicinity of $\mathrm{Pe} \approx 10$.  In Appendix \ref{app:lagpar}, a more detailed discussion of the asymptotic dependencies are provided.

\begin{figure*}[t]
	\begin{center}
\includegraphics[width=\textwidth]{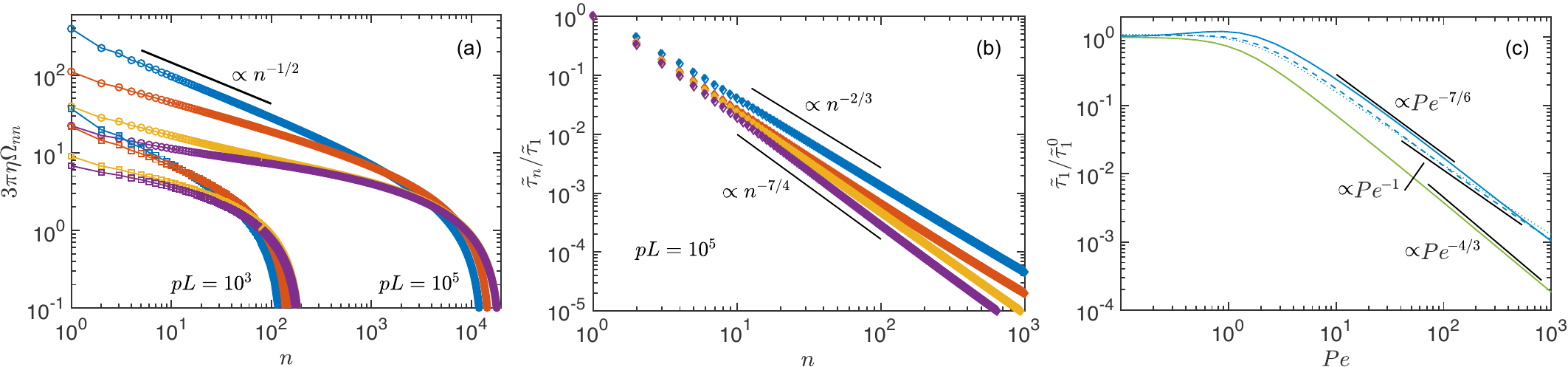}
	\end{center}
	\caption{(a) Mode-number dependence of the Oseen tensor $\Omega_{nn}$ for polymers of length $pL=10^3$ (squares) and $pL=10^5$ (circles) and the P\'eclet numbers $\mathrm{Pe}=10^{-2}$ (blue),  $1$ (orange), $50$ (yellow), and $10^3$ (purple) (top to bottom). 
(b) Mode-number dependence of the relaxation times $\tilde{\tau}_n$ for flexible polymers of length $pL=10^5$ and the P\'eclet numbers $\mathrm{Pe} = 0$ (blue),  $1$ (orange), $50$ (yellow), and $10^3$ (purple) (top to bottom). 
(c) Longest polymer relaxation time $\tilde{\tau}_1$, Eq.~\eqref{eq:tau_asym_flex}, normalized by the corresponding passive value  $\tilde{\tau}_1^0$ as function of the P\'eclet number $\mathrm{Pe}$ for flexible polymers with $pL=50$ (dotted), $1.5 \times 10^2$ (dashed), and $10^3$ (solid blue, top). The bottom solid curve (green) corresponds to  an active polymer in absence of HI for $pL=10^3$, where $\tau_1 \sim \mathrm{Pe}^{-4/3}$. The short  lines (black) indicate power-law dependencies  in the respective regimes.
} \label{fig:hyd_relaxation}
\end{figure*}

Figure~\ref{fig:hyd_relaxation}(a) depicts the dependence of the preaveraged Oseen tensor on the mode number for flexible polymers. For a passive polymer, we obtained the dependence $\Omega_{nn} \sim n^{-1/2}$ of the Zimm model \cite{doi:86} over a range of mode numbers, which depends on $pL$. With increasing P\'eclet number, both  the values of  $\Omega_{nn}$ and the magnitude of the slope decrease substantially. As a consequence, at high P\'eclet numbers, $\Omega_{nn}$ does not contribute to the mode-number dependence of the relaxation time anymore, as is reflected in Fig.~\ref{fig:hyd_relaxation}(b). Zimm-type relaxation times  $\tilde \tau_n \sim n^{-3/2}$ are obtained for the passive polymer (Fig.~\ref{fig:hyd_relaxation}(b)) \cite{harn:96}.  With increasing P\'eclet number, the mode-number  dependence  changes to $\tilde \tau_n \sim n^{-7/4}$ for $\mathrm{Pe}=10^3$, a dependence very close to that of a free-draining, non-hydrodynamic Rouse polymer \cite{doi:86}. This emphasizes the diminishing effect of  hydrodynamic interactions with increasing activity.

The activity-dependence of the longest polymer relaxation time is displayed in Fig~\ref{fig:hyd_relaxation}(c). The decline of  $\tilde \tau_1$ with increasing $\mathrm{Pe}$ is determined by the stretching coefficient $\mu$ and the implicit dependence of $\Omega_{11}$ on $\mu(\mathrm{Pe})$. The shift to larger $\mathrm{Pe}$ of the curves in presence of HI reflects its influence  on the relaxation times, specifically the influence on $\Omega_{11}$. The latter is also responsible for values $\tilde \tau_1/\tilde \tau_1^0 > 1$ ($\mathrm{Pe} \approx 1$), because $\Omega_{11}$ decreases with increasing $\mathrm{Pe}$ (Fig.~\ref{fig:hyd_relaxation}(a)). As discussed in App.~\ref{app:lagpar}, $\mu$ is essentially independent of hydrodynamic interactions for $pL \ll \mathrm{Pe}$, hence the decline  of $\tilde \tau_1$ with increasing $\mathrm{Pe}$ ($\mathrm{Pe} \gg1$) for $pL =50$  is solely determined by $\mu$ and $\tilde \tau_1 \sim 1/\mathrm{Pe}$. Similarly, the asymptotic behavior for $pL=10^3$ is determined by  $\mu$, with $\tilde \tau_1 \sim 1/\mathrm{Pe}^{4/3}$, the dependence of a polymer in absence of hydrodynamic interactions. However, for very flexible polymers, $pL \gtrsim 10^3$, HI gives rise to an intermediate regime, $10< \mathrm{Pe} <10^3$, where $\tilde \tau_1 \sim \mathrm{Pe}^{-7/6}$.  The difference to a decay with $\tilde \tau_1 \sim \mathrm{Pe}^{-1}$ seems subtle, but is essential and strongly affects the conformational and dynamical properties of a polymer,  as will be discussed in Sec.~\ref{sec:conformations} and \ref{sec:cont_dyn}.

\begin{figure}
	\begin{center}
		\includegraphics[width=\columnwidth]{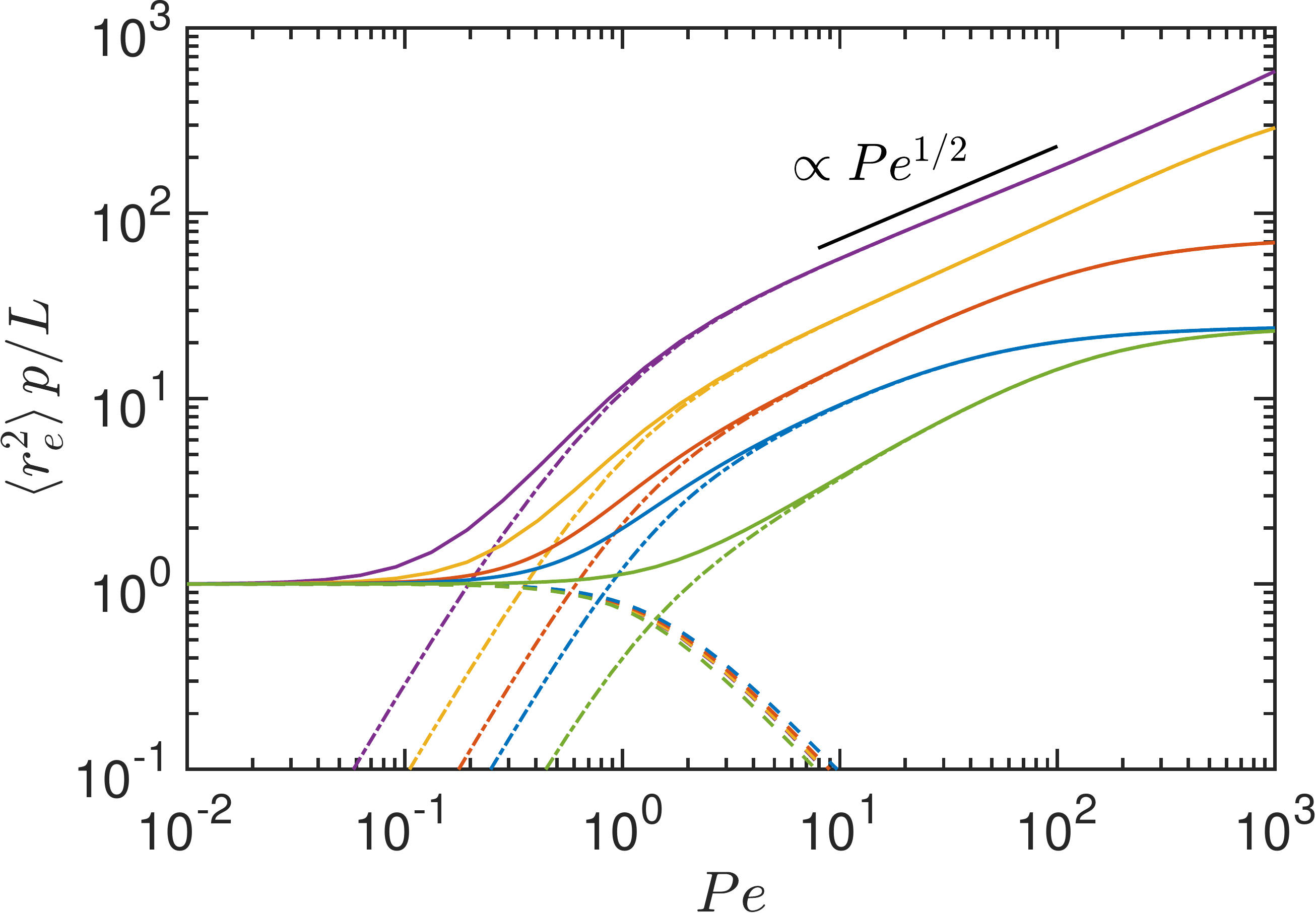}
	\end{center}
	\caption{Polymer mean square end-to-end distance $\left < \bm r ^2 _e\right >$  scaled by the equilibrium value $L/p$ in the presence of HI as a function of the P\'eclet number $\mathrm{Pe}$ for flexible polymers of length $pL = 50$ (blue), $pL = 1.5 \times 10^2$ (orange), $pL = 10^3$ (yellow), and $pL = 10^4$ (purple) (bottom to top). The green curve corresponds to the free-draining flexible polymer with $pL = 50$. The dashed curves represent the passive contribution with the relaxation times $\tau_n$ and the dashed-dotted curves the active part with $v_0^2$ in Eq.~\eqref{eq:end-to-end}. The short  line (black) indicates a power-law dependence in the respective regime.} \label{fig:end_sep_ED_BD}
\end{figure}

\begin{figure}
	\begin{center}
		\includegraphics[width=\columnwidth]{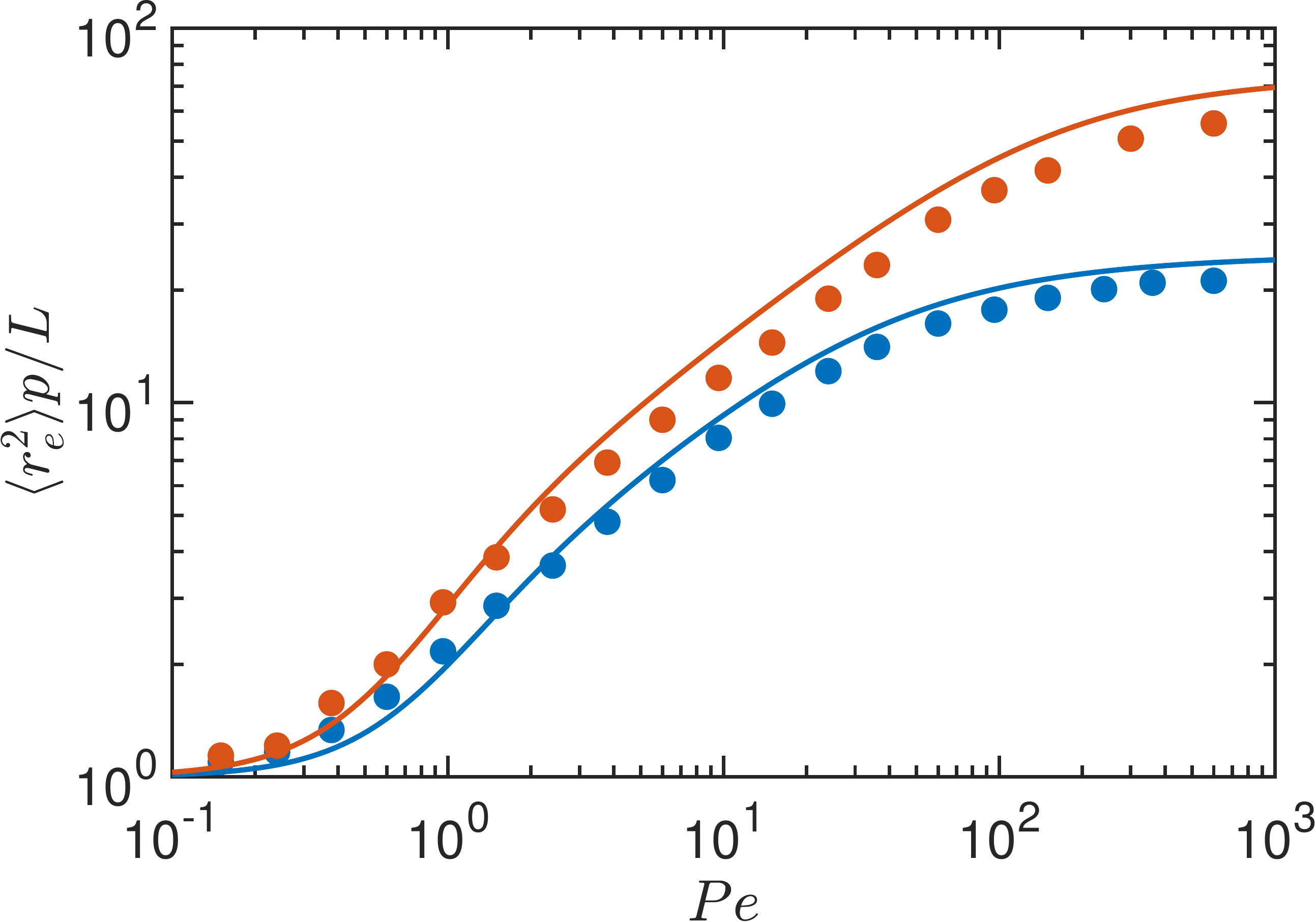}
	\end{center}
	\caption{Comparison of the dependence of  polymer mean square end-to-end-distance  on the P\'eclet number obtained from analytical theory (lines) and BD simulations (bullets) for polymers with $pL = 50$ (blue, bottom) and $pL = 1.5 \times 10^2$ (orange, top).} \label{fig:end_sep_ED_theo_50_150}
\end{figure}

\section{Conformational properties} \label{sec:conformations}

The conformational properties of the polymers are characterized by their mean square end-to-end distance $\langle \bm r_e ^2 \rangle = \langle (\bm r(L/2) - \bm r(-L/2))^2 \rangle$, which is
\begin{align} \label{eq:end-to-end}
\lla \bm r_e^2\rra = \frac{8}{L} \sum_{n, \, \text{odd}} \left( \frac{k_BT \tau_n}{\pi \eta} + \frac{v_0^2l \tau_n^2}{1+\gamma_R \tilde \tau_n} \right)
\end{align}
in terms of the mode amplitudes of Eq.~\eqref{eq:chi_st_st}. Numerical results for $\langle \bm r^2_e \rangle$ are displayed in  Fig.~\ref{fig:end_sep_ED_BD}. As in simulations (cf. Fig.~\ref{fig:end_ED_BD}), polymers swell stronger with increasing activity than free draining active polymers,  and their size saturates  at  $L^2/2$ for $Pe \to \infty$, the value of the free-draining case. Similarly to free-draining polymers or polymers with self-propelled monomers, the thermal contribution, proportional to $k_BT$, decreases  and the active term, proportional to $v_0^2$, increases with increasing $\mathrm{Pe}$. 
However,  the swelling behavior is distinctly different compared to  those two cases, which is reflected by the respective dependence on the relaxation times $\tau_n$ and $\tilde \tau_n$. Comparing the relaxation-time dependence of the active term (with $v_0$) of a free-draining polymer, 
$\tau_n^2/(1+\gamma_R \tau_n)$ \cite{eise:16}, of a polymer with self-propelled monomers,  $\tilde \tau_n^2/(1+\gamma_R \tilde \tau_n)$ \cite{mart:19}, and that of Eq.~\eqref{eq:end-to-end}, we find
\begin{align}
\frac{\tau_n^2}{1+\gamma_R \tilde \tau_n} \ge  \frac{\tau_n^2}{1+\gamma_R  \tau_n} \ge \frac{\tilde \tau_n^2}{1+\gamma_R \tilde \tau_n} ,
\end{align}
because $\tilde \tau_n \le \tau_n$. Hence, the externally-driven polymer swells strongest with increasing P\'eclet number, and swelling sets in at smaller $\mathrm{Pe}$. This is reflected in the shift of the dashed-dotted lines in Fig.~\ref{fig:end_sep_ED_BD} to smaller $\mathrm{Pe}$ with increasing $pL$, whereas respective curves shift to larger $\mathrm{Pe}$ in case of polymers with self-propelled monomers, associated with polymer shrinkage \cite{mart:19}. This reveals the distinct influence of the character of the active noise on the  polymer conformations in presence of hydrodynamic interactions.

The asymptotic limit for $\mathrm{Pe} \to \infty$ can be obtained analytically. The term  $\gamma_R \tilde \tau_n \ll 1$ for $\mathrm{Pe} \to \infty$  (cf. App.~\ref{app:lagpar}) and, thus, can be neglected in Eq.~\eqref{eq:end-to-end}. Evaluation of the sum over modes with the relaxation times \eqref{eq:relax_flex} and insertion of Eq.~\eqref{app:mu_linear} then gives  $\langle \bm r_e^2 \rangle = L^2/2$. This result is in close agreement with simulations, which yield a somewhat smaller value, as shown in Fig.~\ref{fig:end_sep_ED_theo_50_150}. Moreover, the asymptotic limit is identical with that of a free-draining polymer \cite{eise:16}, in contrast to a polymer of self-propelled monomers \cite{mart:19}.

The enhanced swelling of the externally-driven flexible polymer can be understood as follows. In the regime  of strong polymer swelling, e.g., $0.1 < \mathrm{Pe} < 100$ for $pL =50$ in Fig.~\ref{fig:end_sep_ED_BD}, $\gamma_R \tilde \tau_1 \gg 1$ and the active velocity-dependent term in Eq.~\eqref{eq:chi_st_st}  can be approximate  by 
\begin{align}
\frac{v_0^2 l \tau_n }{\gamma_R} \left(1 + 3 \pi \eta \Omega_{nn} \right) ,
\end{align}
which is by the contribution $3 \pi \eta \Omega_{nn}$ larger than the term in absence of HI. Formally, we can introduce an effective larger velocity $v_0 \sqrt{1+3 \pi \eta \Omega_{nn}}$, which corresponds to an effectively higher P\'eclet number and, hence, a stronger polymer swelling.  According to Eq.~\eqref{eq:chi_n}, both  the active velocity $\bm v_n(t)$ and the stochastic force $\bm \varGamma_n(t)$ are enhanced by the hydrodynamic tensor $H_{nn}$. However,  the hydrodynamic effect disappears in the thermal contribution of the correlation function \eqref{eq:chi_st_st}, because of  the fluctuation-dissipation relation  Eq.~\eqref{eq:corr_stoch_mode}. Hence, the strong hydrodynamic effect on polymer conformations is a consequence of the independence of the rotational dynamics  from the translational hydrodynamic tensor (cf. Eq.~\eqref{eq:mode_corr_v}).  

Simulations (Fig.~\ref{fig:end_ED_BD}) and analytical calculations (Fig.~\ref{fig:end_sep_ED_BD}) predict the swelling behavior $\langle \bm r_e^2 \rangle \sim \mathrm{Pe}^{1/2}$ over a range of P\'eclet numbers, where the range increases with increasing $pL$. This dependence on $\mathrm{Pe}$ is markedly different from that of free-draining polymers and those with self-propelled monomers; in the latter case the exponent is larger than unity \cite{mart:19}. This difference rests upon a particular dependence of the dynamics on hydrodynamic interactions, reflected in the $\mathrm{Pe}$ dependence of the relaxation time $\tilde \tau_1$ (Fig.~\ref{fig:hyd_relaxation}(c)). This can be shown analytically. First of all, the mode-number dependence of the relaxation times $\tilde \tau_n$ is well described by a power law, specifically for $pL=10^3$,  $\tilde \tau_n \approx \tilde \tau_1/n^2$ (Fig.~\ref{fig:hyd_relaxation})(b). Second, in the relevant $\mathrm{Pe}$ regime  $\gamma_R \tilde \tau_n \gg1$, hence,  Eq.~\eqref{eq:end-to-end} yields
\begin{align} \label{eq:re_approx}
\lla \bm r_e^2 \rra \sim \frac{\mathrm{Pe}^2}{\mu^2 \tilde \tau_1} \sim \sqrt{\mathrm{Pe}} ,
\end{align}
with $\mu \sim \mathrm{Pe}^{4/3}$ (Eq.~\ref{app:mu_nonlinear}) and $\tilde \tau_1 \sim \mathrm{Pe}^{-7/6}$ (Fig.~\ref{fig:hyd_relaxation}(c)), relations appropriate for $pL=10^3$. It is the $\mathrm{Pe}$ dependence of the relaxation time $\tilde \tau_1$  which is decisive for the relation \eqref{eq:re_approx}. In the absence of HI, $\tau \sim 1/\mu \sim \mathrm{Pe}^{-4/3}$ and $\langle \bm r_e^2 \rangle \sim \mathrm{Pe}^{2/3}$ \cite{eise:16}, which is a substantially stronger $\mathrm{Pe}$ dependence. The seemingly rather small difference between the exponent   $-4/3=-8/6$, valid in absence of HI,  and $-7/6$, valid with HI,  of the  relaxation time is decisive and leads to a weaker swelling of the externally driven polymer with increasing $\mathrm{Pe}$.

The theoretical approach very well reproduces the simulation data, as shown in Fig.~\ref{fig:end_sep_ED_theo_50_150}. The analytical theory somewhat overestimates the asymptotic value as a consequence of the mean-field-type constraint for the bond length (Eq.~\eqref{eq:constraint_general}).

We like to emphasize that the swelling of active polymers is determined by their  inextensibility, as is evident from the results of this section. Only by taking this polymer feature suitably into account, e.g., via the constraint \eqref{eq:constraint_general}, the qualitative correct behavior is obtained theoretically \cite{eise:16,eise:17,mart:18.1,mart:19,mous:19}. Approaches neglecting such a condition predict swelling, which qualitatively and quantitatively disagrees with simulation results.

\section{Dynamical properties} \label{sec:cont_dyn}

The polymer dynamics is analyzed  in terms of the monomer mean square displacement (MSD) averaged over the polymer contour
\begin{align}  \label{eq:def_msd} \nonumber
 \lbar {\lla  \Delta \bm r^2(t) \rra} = & \ \frac{1}{L} \int \lla  (\bm r(s,t) - \bm r(s,0))^2 \rra ds  \\
  = & \ \lla \Delta  \bm r_{cm}^2(t) \rra +  \lla \Delta  \bm r_{0}^2 (t) \rra +  \lla \Delta  \bm r_{a}^2(t) \rra ,
\end{align}
with the center-of-mass mean square displacement
\begin{align} \label{eq:amp_CMmsd_SP} \nonumber
&\lla \Delta  \bm r_{cm}^2(t) \rra  =  \ H_{00} \frac{6k_BT }{L}  t \\ & \hspace*{1.2cm} + (1+ 3 \pi \eta \Omega_{00}) \frac{2 v_0^2 l }{\gamma_R^2 L} \left( \gamma_R t -1 + e^{-\gamma_R t} \right) ,
\end{align}
$H_{00}=(1+3\pi \eta \Omega_{00})/(3\pi \eta)$, the activity-modified equilibrium-like internal dynamics contribution
\begin{align}  \label{eq:msd_eq}
\lla  \Delta  \bm r^2_0(t) \rra  =  \frac{1}{L}\sum_{n=1}^{\infty}   \frac{2 k_BT \tau_n}{\pi \eta} \left(1 - e^{-t/\ttau_n}\right)  \ ,
\end{align}
and the active contribution
\begin{align} \label{eq:msd_ac}
\lla  \Delta  \bm r^2_a(t) \rra  =  \frac{1}{L}\sum_{n=1}^{\infty}  \frac{2 v_0^2 l \tau_n ^2}{1+ \gamma_R \ttau_n } \left( 1 - \frac{e^{-\gamma_R t} - \gamma_R \ttau_n e^{-t/\ttau_n}  }{1 - \gamma_R \ttau_n}  \right) .
\end{align}

The passive parts of $\lbar {\langle  \Delta \bm r^2(t) \rangle} $---in $\langle \Delta  \bm r_{cm}^2(t) \rangle$  and $\langle  \Delta  \bm r^2_0(t) \rangle $---,
are, aside of the $\mu$-dependence of the relaxation times, identical with the dynamics of the Zimm model, or that of a semiflexible polymer in presence of HI \cite{doi:86,petr:06,mart:19}.

The center-of-mass MSD exhibits the same time-dependent terms as an active polymer without HI and a polymer with self-propelled monomers. For $t \to \infty$, $\langle \Delta  \bm r_{cm}^2(t) \rangle$ dominates the total MSD, increasing linearly in time with the diffusion coefficient
\begin{align} \label{eq:effective_diff_coef}
D = \frac{1+3 \pi \eta \Omega_{00}}{L}\left( \frac{k_BT}{3 \pi \eta}  + \frac{v_0^2 l }{3\gamma_R } \right)  ,
\end{align}
which is the diffusion coefficient in absence of HI, term in brackets, modified by hydrodynamics, $\Omega_{00}$; the latter  depends on polymer length and P\'eclet number.  Figure~\ref{fig:hyd_relaxation}(a) indicates   a substantial increase of $\Omega_{00}$ with polymer length, a  decrease with increasing $\mathrm{Pe}$, and $\Omega_{00}$ seems to approach a $\mathrm{Pe}$-independent value for $\mathrm{Pe} \gg 1$.

The site-averaged MSD  in the center-of-mass reference frame,  $\langle \Delta  \bm r_{0}^2 (t) \rangle +  \langle \Delta  \bm r_{a}^2(t) \rangle$, exhibits three distinct regimes:

\begin{figure}[t]
	\begin{center}
		\includegraphics[width=\columnwidth]{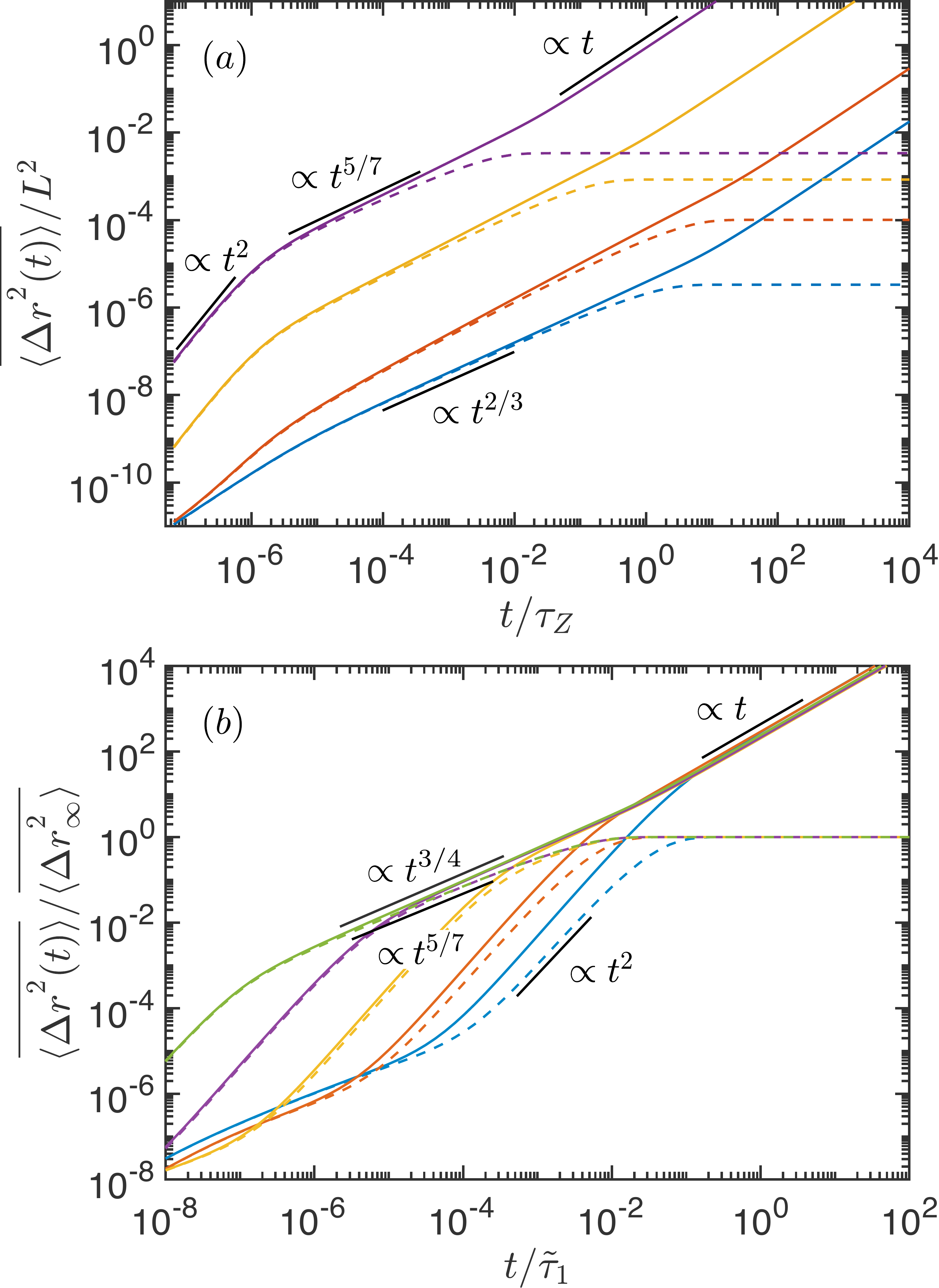}
	\end{center}\caption{(a) Mean square displacement of flexible polymers for $pL=10^5$ and the P\'eclet numbers $\mathrm{Pe}= 0$ (blue),  $1$ (orange), $50$ (yellow), and $10^3$ (purple) (bottom to top). The Zimm relaxation time  $\tau_Z = \eta (L/p)^{3/2} /(\sqrt{3\pi} k_B T)$  is the longest relaxation time of the passive polymer. (b) The P\'eclet number is $\mathrm{Pe}=115$ and  $pL=50$ (blue), $1.5 \times 10^2$ (orange), $10^3$ (yellow), $10^4$ (purple), and $10^5$ (green) (bottom to top). The dashed lines correspond to the MSD in the polymer center-of-mass reference frame, and the solid lines are the overall MSD. The short  lines (black) indicate  power-laws in the respective regimes.} \label{fig:msd_ED_theory_hyd}
\end{figure}

\begin{itemize}[leftmargin=*]
\item $t \to 0$ --- The MSD is dominated by Eq.~\eqref{eq:msd_eq}, and all modes contribute. With $\tau_n = \tau_R /(\mu n^2)$ for a flexible polymer,  conversion of the sum to an integral yields
\begin{align} \label{eq:msd_equil_approx}
\lla  \Delta  \bm r^2_0(t) \rra  =  \frac{2L}{\pi^2 p \mu} \left(\frac{t}{\ttau_1} \right)^{2/3} \int_0^{\infty} dx \frac{1-e^{-x^{3/2}}}{x^2} .
\end{align}
This is the same relation as obtained for a passive system, except that $\mu$ and $\ttone$ depend on activity, and a polymer with self-propelled monomers \cite{mart:19}.
\item $t/\tilde \tau_1$ and  $\gamma_R t \ll   1 $ --- Taylor expansion of the exponential functions in Eq.~\eqref{eq:msd_ac} yields
\begin{align} \label{eq:msd_ballistic}
\lla  \Delta  \bm r^2_a(t) \rra  =  \frac{v_0^2 l \gamma_R}{L} \sum_{n=1}^{\infty} \frac{\tau_n^2}{\ttau_n(1+ \gamma_R \ttau_n) } t^2 ,
\end{align}
consistent with the observed ballistic regime in Fig.~\ref{fig:msd_ED_theory_hyd}. This  regime and its dependence on activity and polymer properties is in qualitative agreement with the simulation results of Fig.~\ref{fig:msd_sim}.
\item $1/\gamma_R \ll t \ll \tilde \tau_1$ --- With $\gamma_R \tilde \tau_1 \gg1 $,  the MSD is given by
\begin{align} \label{eq:msd_segm}
\lla  \Delta  \bm r^2_a (t) \rra  =   \frac{2 v_0^2 l}{\gamma_R L}\sum_{n=1}^{\infty}  \frac{\tau_n^2}{\ttau_n}  \left( 1 - e^{-t/\ttau_n}  \right)  .
\end{align}
The relaxation times $\tilde \tau_n$  are well described by the power-law $\tilde \tau_n = \ttone/n^{\gamma}$ (cf. Fig.~\ref{fig:hyd_relaxation}). Inserting this relation and replacing the sum by an integral, Eq.~\eqref{eq:msd_segm} yields
\begin{align}  \label{eq:msd_active}
\lla \Delta  \bm r^2_a (t) \rra  =
 \frac{2 v_0^2 l \tau_R^2}{\mu^2 \gamma_R L}  \left(\frac{t}{\ttone} \right)^{\gamma'}  \int_0^{\infty} dx \ \frac{1-e^{ -x^{\gamma}}}{x^{4-\gamma}}  ,
\end{align}
with $\gamma' = {3/\gamma-1}$. 
For $\mathrm{Pe}>50$ and $pL = 10^3$, the power-law exponent is  close to $\gamma = 7/4$, hence,
\begin{align} \label{eq:power_law}
\lla \Delta  \bm r^2_a(t) \rra \sim  t^{5/7} .
\end{align}
This time dependence is in close agreement with the numerical result displayed in Fig.~\ref{fig:msd_ED_theory_hyd}. By the interplay between activity and hydrodynamic interactions, a new power-law regime emerges for the inter-molecular MSD. The seemingly small difference between the exponent of the relaxation times  $\gamma = 7/4$ and the value $\gamma = 2$  for a Rouse polymer, implies a significantly different power-law of the  MSD, namely an exponent $\gamma' =5/7$ for the current active polymer vs. $\gamma' = 1/2$ for a Rouse polymer \cite{eise:17}. Moreover, the type of active force matters---calculations for self-propelled monomers yield the exponent $\gamma' =2/5$ \cite{mart:19}, which is even smaller than the value for  free-draining polymers
This emphasizes the strong and dominating influence of hydrodynamic interactions on the dynamics of active polymers.

The overall monomer MSD \eqref{eq:def_msd} exhibits even a different power-law regime $\lbar {\langle  \Delta \bm r^2(t) \rangle} \sim t^{3/4}$ for $pL \gtrsim 10^3$, by an additional contribution of the center-of-mass MSD. Evidently, a splitting of the  center-of-mass-site MSD, $\langle \Delta  \bm r_{0}^2 (t) \rangle +  \langle \Delta  \bm r_{a}^2(t) \rangle$, from the overall MSD is  not possible, even for very long polymers.
\end{itemize}

\begin{figure}[t]
	\begin{center}
		\includegraphics[width=\columnwidth]{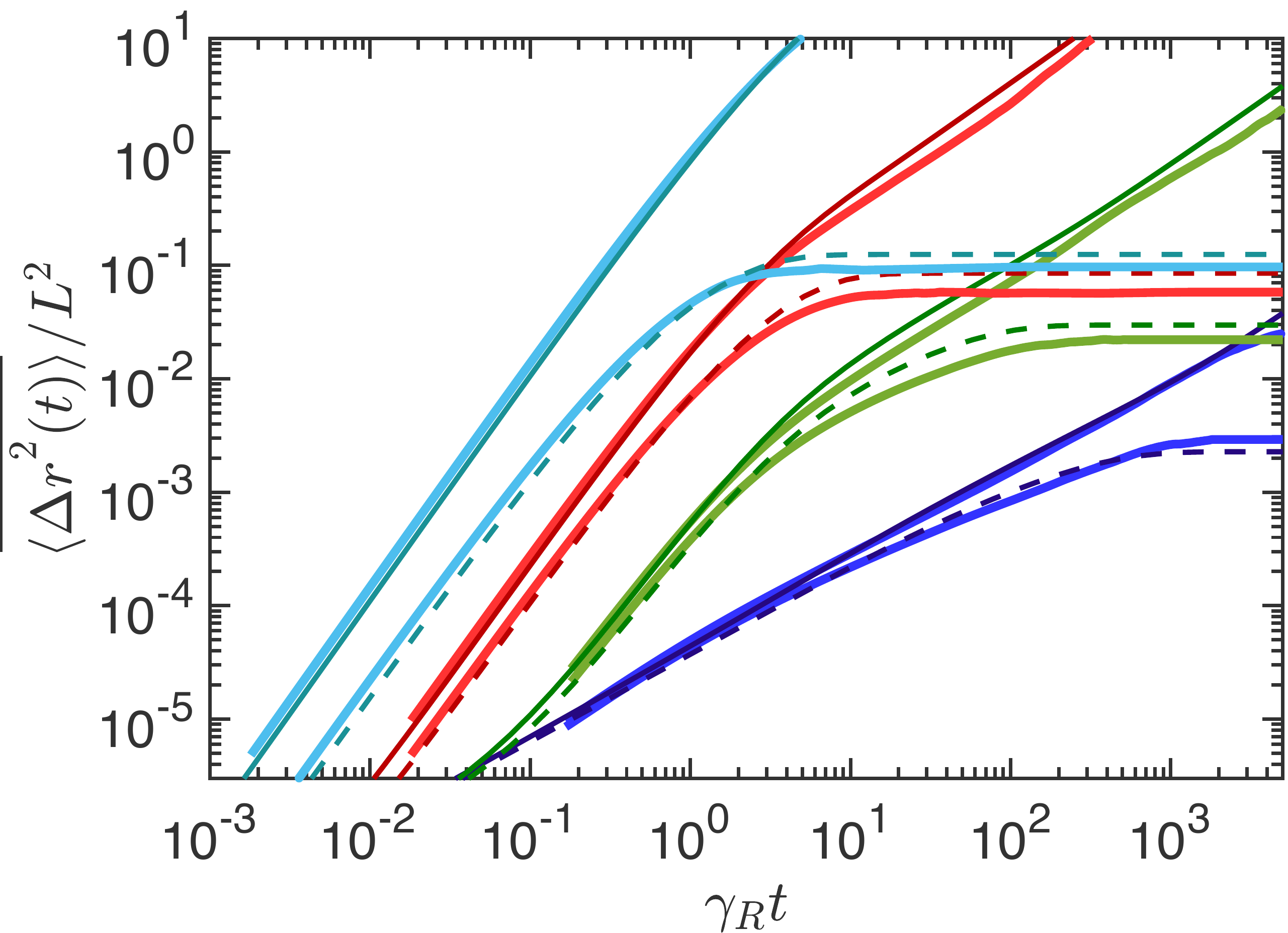}
	\end{center}\caption{Comparison of the mean square displacement  obtained in simulations (broad solid lines; Fig.~\ref{fig:msd_sim}) with analytical theory (thin solid and dashed lines; Eq.~\eqref{eq:def_msd}) for the P\'eclet numbers $\mathrm{Pe} = 0$ (blue), $10^1$ (green), $10^2$ (red), and $10^3$ (cyan) (bottom to top). The monomer number is $N_m=150$ and $pL=L/l=150$, respectively.  The dashed lines and the respective broad solid lines correspond to the MSD in the polymer center-of-mass reference frame.} \label{fig:msd_theo_sim_150}
\end{figure}

\rev{Figure \ref{fig:msd_theo_sim_150} presents a comparison of the mean square displacements of the discrete polymer of Fig.~\ref{fig:msd_sim} with analytically results. The results agree very well considering the limited statistical accuracy in the simulation results, the approximation in the analytical evaluation of the hydrodynamic tensor, and the overestimation of the extension of active polymers for $\mathrm{Pe}\gg 1$ (cf. Fig.~\ref{fig:end_sep_ED_theo_50_150}) as a consequence of the mean-field-type constraint for the bond length. The latter is reflected by the analytical results exceeding the asymptotic values of the MSD in the polymer center-of-mass reference frame for $t \to \infty$, which is theoretically twice the radius of gyration, but somewhat smaller in simulations. }

\section{Summary and Conclusions} \label{sec:summary}

We have studied the conformational and dynamical properties of semiflexible active polymers in presence of hydrodynamic interactions by simulations and analytical theory. In the simulations, we consider the overdamped dynamics of a bead-spring polymer, including hydrodynamic interactions via the Rotne-Prager-Yamakawa hydrodynamic tensor. Moreover, we present an implementation of the active polymer in the multiparticle collision dynamics approach. Comparison of the polymer conformational properties at various P\'eclet numbers and polymer stiffness yields quantitative agreement between simulations employing the hydrodynamic tensor and the MPC method, respectively. The MPC approach opens possibilities to study active polymers in situations, where a tensor description is extremely difficult and demanding, as for polymers confined in channels.
In the analytical treatment, the Gaussian semiflexible polymer model is adopted, taking into account the polymer inextensibility in a mean-field manner by a constraint for the average contour length. Here, hydrodynamic interactions are taken into account by the preaveraged Oseen tensor.
In any case, activity is modeled as a Gaussian colored noise process with an exponential temporal correlation. This activity is assumed to be imposed externally onto the monomers by the embbeding active bath. As a consequence, the active force gives rise to monomer Stokeslet flow fields, in contrast to self-propelled monomers which are active force free \cite{mart:19}.  Further Stokeslets appear by intramolecular  forces due to bond, bending, volume exclusion, and thermal forces. 
 
Our studies reveal a strong effect of hydrodynamics on both conformations and dynamics. As a consequence of the activity-induced Stokeslets, polymers swell monotonically and stronger with increasing P\'eclet number than active polymers in absence of hydrodynamic interactions  \cite{eise:16} and  polymers composed of self-propelled monomers \cite{mart:19}. In the asymptotic limit of an infinite P\'eclet number, the same finite mean square end-to-end  distance is assumed as for a free-draining active polymer, a value which is large than that of  polymers with self-propelled monomers. As we have shown by analytical calculations, in this limit hydrodynamic interactions become irrelevant. Moreover, we find a broad range of P\'eclet numbers, where the mean square end-to-end distance increase as $\mathrm{Pe}^{1/2}$ for a wide range of stiffnesses. This increase is slower compared to that of the other two types of active polymers. Here,  the dependence of the longest relaxation time on the P\'eclet number plays a decisive role, with $\tilde \tau_1$ being strongly affected by hydrodynamic interactions. 

Qualitatively, we explain the enhanced polymer swelling with increasing $\mathrm{Pe}$ by a hydrodynamically accelerated active velocity. In turn, this implies an apparent  higher P\'eclet number, followed by stronger swelling.  The flow field induced by translating parts of the polymer advects  monomers/sites and leads to an accelerated dynamics.  
Similarly, an enhanced thermal force appears in the solution of the normal-mode amplitudes, $\bm \chi_n$, however, this effect is compensated by the fluctuation-dissipation relation. This implies that the thermal parts of conformational quantities at equilibrium  are explicitly independent of hydrodynamic interactions.    This does not apply to the active velocity, because its  temporal correlation function is independent of  hydrodynamics.  

The polymer dynamics is determined by two relaxation processes, the orientational relaxation of an active site/monomer, and the polymer internal relaxation modes. This is reflected in distinct time regimes in the polymer mean square displacement. At short times $t/\ttone \ll 1$ and $\gamma_Rt \ll 1$, activity implies to a ballistic regime, with an enhanced dynamics compared to a passive polymer. For $1/\gamma_R \ll t \ll \ttone$, the MSD is dominated by the internal dynamics, and a polymer-characteristic subdiffusive regime appears. Again, activity and hydrodynamics play a decisive role, leading to a power-law dependence of the site MSD in the polymer center-of-mass reference frame with an exponent $\gamma'=5/7$, larger  than that of a free draining  and an active polymer with self-propelled monomers. In the asymptotic limit of long times, the free-draining active diffusive coefficient is amplified by hydrodynamics, in the same way as the thermal diffusion coefficient (cf. Eq.~\eqref{eq:effective_diff_coef}).

The analytical calculations and the good agreement with simulations indicate that a suitable account of the fixed polymer contour length is  essential for a qualitative correct description of the active polymer conformations. In the analytical calculations, we have been taking this constraint into account in a mean-filed manner by the average mean square contour length (Eq.~\eqref{eq:constraint_general}). This leads to a strong activity dependence of the relaxation times and consequently the observed increase of the mean square end-to-end distance, $\langle \bm r_e^2 \rangle  \sim  \mathrm{Pe}^{1/2}$, and a saturation of $\langle \bm r_e^2 \rangle$  for $\mathrm{Pe} \to \infty$. Omission of this polymer property, as  common in the theoretical description of active polymers by the  Rouse/Zimm model, leads to artifacts  especially at moderate and large activities.

In conclusion, in presence of hydrodynamics, the properties of active polymers consisting either of self-propelled monomers or experiencing an external driving force with the same temporal correlation function  are substantially different. In the first case, even flexible polymers shrink at moderate  P\'eclet numbers and swell for larger $\mathrm{Pe}$; in the second case, polymers swell monotonically for all $\mathrm{Pe}$, and the polymer size is (siginificant)  larger for all P\'eclet numbers. The difference in the coupling to the flow field  leads to a reduced or enhanced active velocity, and is reflected in the polymer conformations and dynamics. 

Experimentally, an externally driven polymer can in principle be realized by forcing a chain of colloidal particles  by optical tweezers \cite{grie:03}. Optical forces are very well suited to manipulate objects as small as  $5$ nm and up to hundreds of micrometers \cite{grie:03}. Combined with  computer-generated holograms, many particles can be manipulated with a single laser beam at the same time. The example of an optical pump of Ref.~\cite{terr:02} illustrates the possibility to manipulate several colloidal particle simultaneously. Such a setup is therefor well suited to  actuated a colloidal polymer \cite{loew:18}. The persistent colloid motion can be controlled by the tweezer light field which translates them in random directions with an exponential temporal orientation correlation function. \\

\section*{Acknowledgements}

This research was funded by the European Union’s Horizon 2020 research and innovation programme under Grant agreement No. 674979-NANOTRANS. Financial support by the Deutsche Forschungsgemeinschaft (DFG) within the priority program SPP 1726 ‘‘Microswimmers—from Single Particle Motion to Collective Behaviour’’ is gratefully acknowledged. Moreover, the authors gratefully acknowledge the computing time granted through JARA-HPC on the supercomputer JURECA at Forschungszentrum J\"ulich.

\begin{appendix}

\section{Asymptotic stretching coefficient} \label{app:lagpar}

The active contribution in the Eq.~\eqref{eq:constraint} can by written as
\begin{align}
&\sum_{i=1}^{\infty} \frac{v_0^2l \tau_n^2}{1+\gamma_R \tilde \tau_n} \zeta_n^2 =
\frac{\mathrm{Pe}^2 pL^2}{9 \mu^2 \Delta^2 \pi^2 } \\ \nonumber &
\hspace*{1cm} \times \sum_{i=1}^{\infty} \left[n^2 + \frac{2 (pL)^2}{3 \mu \Delta \pi^2 (1+ 3 \pi \eta \Omega_{nn})} \right]^{-1} ,
\end{align}
when we set $d_H \equiv l$. The stretching coefficient, $\mu$, increasing with increasing $\mathrm{Pe}$. Hence, for $(pL)^2 \ll \mu$, the second term in the brackets can be neglected. Then, we obtain from Eq.~\eqref{eq:constraint}
\begin{align} \label{app:mu_linear}
\mu = \sqrt{\frac{pL}{6}} \frac{\mathrm{Pe}}{3 \Delta} .
\end{align}
In the opposite limit, $pL \gg \mathrm{Pe}$, the sum over $n$ is dominated by the second term in the bracket for small $n$, and the mode-number dependence is determined by the preaveraged Oseen tensor. With increasing P\'eclet number, higher modes become important, at the same time $\Omega_{nn}$ becomes less relevant. Neglecting the hydrodynamic contribution, or at least its mode-number dependence, the sum over modes can be evaluated, and Eq.~\eqref{eq:constraint} yields \cite{eise:16}
\begin{align} \label{app:mu_nonlinear}
\mu \sim \mathrm{Pe}^{4/3} .
\end{align}

\end{appendix}



\end{document}